\ifdefined\pdftexversion\pdfoutput=1\fi
\documentclass[11pt]{article}

\usepackage{iftex}
\ifPDFTeX
  \usepackage[T1]{fontenc}
  \usepackage[utf8]{inputenc}
  \usepackage{lmodern}
  \usepackage{textcomp}
  \DeclareUnicodeCharacter{2212}{\ensuremath{-}}
  \DeclareUnicodeCharacter{03B1}{\ensuremath{\alpha}}
  \DeclareUnicodeCharacter{2265}{\ensuremath{\geq}}
  \DeclareUnicodeCharacter{00D7}{\ensuremath{\times}}
  \DeclareUnicodeCharacter{2192}{\ensuremath{\rightarrow}}
  \DeclareUnicodeCharacter{00B6}{\P}
  \DeclareUnicodeCharacter{03C0}{\ensuremath{\pi}}
  \DeclareUnicodeCharacter{03BA}{\ensuremath{\kappa}}
\else
  \usepackage{fontspec}
  \usepackage{newunicodechar}
  \newunicodechar{−}{\ensuremath{-}}
  \newunicodechar{α}{\ensuremath{\alpha}}
  \newunicodechar{≥}{\ensuremath{\geq}}
  \newunicodechar{×}{\ensuremath{\times}}
  \newunicodechar{→}{\ensuremath{\rightarrow}}
  \newunicodechar{π}{\ensuremath{\pi}}
  \newunicodechar{κ}{\ensuremath{\kappa}}
\fi
\usepackage[margin=1in]{geometry}
\usepackage{setspace}
\usepackage{microtype}
\usepackage{graphicx}
\usepackage{booktabs}
\usepackage{tabularx}
\usepackage{array}
\usepackage{caption}
\usepackage{amsmath}
\usepackage[round,authoryear]{natbib}
\usepackage[hidelinks]{hyperref}
\usepackage{xurl}

\makeatletter
\renewcommand\@seccntformat[1]{\csname the#1\endcsname\ifx\relax#1\relax\fi\csname sep@#1\endcsname\quad}
\newcommand\sep@section{.}
\newcommand\sep@subsection{}
\newcommand\sep@subsubsection{}
\makeatother
\newcolumntype{Y}{>{\raggedright\arraybackslash}X}
\newcommand{\figfile}[2]{%
  \IfFileExists{figures/#1}{\includegraphics[width=\linewidth,height=0.74\textheight,keepaspectratio]{figures/#1}}%
  {\fbox{\parbox[c][#2][c]{0.95\linewidth}{\centering\small Figure file \texttt{\detokenize{figures/#1}} not yet drawn.}}}}

\title{Still funded, no longer counted: how NIH's 2025 award reviews changed what the government counts as minority health research}

  \author{Fangfang Xie\textsuperscript{1}, Jingwen Zhang\textsuperscript{2} and Haining Wang\textsuperscript{3}\thanks{Corresponding author. Indiana University School of Medicine, 410 W 10th St, HITS Suite 3000, Indianapolis, IN 46202, USA. Email: \href{mailto:hw56@iu.edu}{hw56@iu.edu}. ORCID: 0000-0002-1196-0918.}\\[4pt]
  \small \textsuperscript{1}Nanjing University\\
  \small \textsuperscript{2}Hubei University\\
  \small \textsuperscript{3}Indiana University School of Medicine}

\date{}

\begin{document}
\maketitle

\begin{abstract}
\noindent Funders know their portfolios through software that classifies award text. The US
National Institutes of Health (NIH) reports its spending in more than 300 categories mined this way, and
work on classification and indicators treats the text as the applicant's to write. In 2025 NIH required ``DEI
language'' removed from awards not supporting DEI activities, and so policed the words it also counts.
We followed population names through 37,790 continuing awards, checking them against practice records. Names were informative: in new awards with trial baselines, a title naming Black populations
predicted a 60-percentage-point higher enrolled share. Text features predicted which names survived, and practice
records added little. NIH's minority health category followed the names: of continuing projects carrying it, 99.4\% with unchanged summaries kept it, against 15.7\% of those whose summaries no longer named a
racial or ethnic population, while the projects kept their funding. Blinded reviewers judged 42 of a random 50 such
losses to have met its definition in FY2024. Comparing both summaries of 70 name losses, they found aims concerning the
population recast in 49. Read alone, 63 FY2025 summaries no longer met the definition. Awards naming sexual and gender
minorities were recorded as terminated 37.8 points more often after adjustment for listed terms, institute and activity. We call the mechanism
word targeting, its boundary population targeting, and its product uncounted science: funded research the count no
longer records. The count followed the edited text, and the edited record cannot say whether the research changed with it.

\medskip
\noindent\textbf{Keywords:} research evaluation; text-based indicators; research funding; NIH; spending categories;
constitutive effects; research classification; research governance; health equity research
\end{abstract}

\section{Introduction}

Research funders increasingly know their own portfolios through software that classifies the text of awards. The US
National Institutes of Health (NIH) has done so since 2008. Its Research, Condition, and Disease Categorization
(RCDC) system mines the titles, abstracts, public health relevance statements and specific aims of awards, and each
year NIH reports to Congress and the public how much it spent in more than 300 categories, from cancer to minority
health \citep{johnson2008,rcdc}. Other funders classify awards by health category and research activity
\citep{hrcs}, and a growing body of work maps research to the Sustainable Development Goals \citep{armitage2020}.
The totals are read as the record of what science the public pays for. Researchers use NIH's
categories to ask whether funding follows disease burden \citep{sampat2013,nguyen2024}, advocates use them to argue
for more \citep{torrey2020}, and NIH staff use them to find gaps in the agency's own portfolio
\citep{brown2022,elena2023}.

Two literatures bear on such records, and both have mostly treated the classified text as the applicant's to write. Work on research classification asks
whether a classification is accurate: whether independent approaches agree \citep{armitage2020}, and whether a
category captures what it names \citep{villani2018,forrest2026}. Work on indicators asks how measurement changes what
is measured: the measured adjust their conduct to the measure \citep{campbell1979,espeland2007}, and indicators come
to define what they were meant to describe \citep{dahlerlarsen2014,derijcke2016}. In both, applicants write the text
for reviewers and software then reads it. We know of no study that follows what happens when the funder itself governs the
text its software reads. The validity of the classification then stops being a fixed property of the classifier. It depends on who
may change the text, and why.

In 2025 NIH governed that text directly. From March 2025, NIH staff guidance directed program officials to review
awards for content related to diversity, equity and inclusion (DEI) before issuing funds. For awards that ``do not
support DEI activities, but may contain language related to DEI'', officials were to request an updated application
or progress report with ``the DEI language removed'' \citep[complaint ¶¶114--116, quoting the guidance of March 4,
2025]{salles2026}. A federal court vacated the directives on June 16, 2025 \citep{apha2025}, and NIH later described
renegotiated titles and abstracts as terms of award \citep{notod26007}.

Prior work on the 2025 events shows that terms of the kind on pre-2025 word lists fell in NIH abstracts from late
2024 \citep{mehta2025} and that terminations fell unequally on topics and investigators
\citep{liu2025,oliveira2026,fieldingmiller2026}. In 2003, investigators whose grants were questioned in Congress
removed words their critics might search for, and the removal made keyword searches unable to show whether funding
had fallen \citep{kempner2008}. This work leaves open what the edits did to the funder's count. External
validation of RCDC itself is sparse and category-specific \citep{nasem2025,villani2018,forrest2026}.

In fiscal year (FY) 2025 the category assignments of a set of continuing projects changed in a way we can trace to their text.
Among continuing projects carrying NIH's category for Racial and Ethnic Minority Health Research, the category was
kept by 99.4\% of those whose summary was unchanged and by 15.7\% of those whose summary no longer named a racial or
ethnic population. Among projects whose practice records show that they study such a population, it was kept by
98.5\% of unchanged summaries, 92.1\% of summaries revised but still naming the population, and 15.0\% of summaries
that no longer named it. The projects were still funded, and their practice records from before 2025 show that they involved the population.

This article follows what the reviews did to that count. We traced population names through 37,790 continuing
awards and checked them against practice records the summaries do not produce: whom linked trials enrolled, how
linked articles were indexed, and whom trial eligibility named. The names were informative. Text features predicted which
names survived, and practice records added little predictive information beyond them. NIH's category followed the names. Of 20 continuing projects with strict practice
evidence whose summaries lost the name, 17 lost the category. Two blinded reviewers judged all 20 to meet the
category's own definition in FY2024. Comparing both summaries of these and of 50 randomly drawn losses, reviewers
found aims concerning the population recast in 49 of 70, and 63 FY2025 summaries that, read alone, no longer met the
definition. The category read the edited text correctly. It no longer shows research that met the category's definition when judged from the project's FY2024 summary and practice records. Where a population itself appeared to be the target, the action moved from the text to the award: awards naming it were terminated far more often.

We argue that policing research through the words a government also counts rewrites the government's record of what science it pays for. The projects continued to be funded. For some of them, trials still listed as
active are the evidence that work on the population goes on. The edited record can no longer say whether the research stayed the same. We call the mechanism
\emph{word targeting}, its boundary \emph{population targeting} and its product \emph{uncounted science}: research the funder still funds, that
pre-2025 practice records show involves the population, and that its own count no longer records.

The article makes three contributions to research evaluation. First, it documents a route by which a text-based indicator can fail without reactivity or gaming. When a condition of funding changes
the indicator's input, the indicator moves with it and records the change as a change in the science. Second, it shows why checks of a classification against
the text it reads cannot see this failure. The category stays consistent with its input, and the error appears as a
break in a series that downstream users read as change. An audit against independent evidence can find a
divergence between category and records in a single year, but it cannot show when or how it arose. Third, it offers a design that can show when and how it arose: follow
the same projects across versions of their text, and compare the category with records made for other purposes.

\section{When the funder edits the text it counts from}
\label{sec:theory}

\subsection{A record edited while the funding continued}

Governments know science through records they compile from it, and act on what the records show. Classifications
make some work visible and leave other work out of view, and they hold only
as long as the practices that feed them hold \citep{bowker1999}. Published counts earn trust by being mechanical,
produced by rules rather than by judgment \citep{porter1995}. RCDC is mechanical in this sense: software assigns the
category from the text, and the totals go to Congress and the public each year \citep{rcdc}. The same property makes
it a channel. Whatever changes the text changes the count, without anyone deciding that it should.
\citet{scott1998} describes states that remade the world to fit their maps. In 2025 the order ran the other way: the
map was edited, and the awards it described kept their funding. That was possible because audit checks documents about an activity rather than the activity \citep{power1997}, and because a published count travels
without the conditions that produced it \citep{espeland1998}. Work on indicators asks how measures change what they
measure \citep{campbell1979,strathern1997,dahlerlarsen2014}, usually through the measured, who adjust their conduct
to the measure \citep{espeland2007}. Here the change came from the funder itself, acting as the authority that releases funds. It requested the change for reasons
unrelated to the measure, and the measure recorded it.

A grant summary carries two kinds of words that a lexicon cannot tell apart. Grant writing is a learned, strategic
genre \citep{serranovelarde2018}. Funded NIH abstracts have grown more promotional over decades \citep{millar2022},
and the broader-impacts criterion of the US National Science Foundation (NSF) produced formulaic statements about
participation \citep{roberts2009,watts2015}. Diversity talk in particular became a broad organizational vocabulary
loosely tied to practice \citep{bell2007,berrey2015}: ceremonial language, in the institutional sense
\citep{meyer1977}. At the same time, NIH's inclusion policy made naming demographic subgroups a standard part of study
design and enrollment reporting \citep{epstein2007}. A summary thus holds \emph{era vocabulary}, words whose spread
suggests that they mark when and for whom a proposal was written, though they can also state substantive aims. Beside
it sit population names, which can record whom a study enrolls. The analysis tests whether
the names are informative and leaves open whether the era vocabulary is ceremonial.

Our own evidence is documentary too. We do not observe research practice. We compare the summary with records made
for other purposes, on other occasions and under other rules: trial registrations, enrollment counts posted to a
registry, and subject indexing assigned before indexing was automated. These records are also documents, but the 2025
review did not govern them. A
change in the summary that these records do not share is a change in the account, whereas a change they share is at
least a candidate change in the research. Their independence is partial: investigators report registrations and
enrollment, registries update on their own schedules, and such records exist only for projects with trials or indexed
articles. Three definitions of the research are therefore in play. Software applies NIH's category definition to the
text. Our practice-record standard captures involvement, and blinded reviewers applied the category
definition themselves. All three are accounts of the research, and our claims concern where they diverge.

\subsection{Word targeting}

We call a text that a funder both acts on and counts from a \emph{shared text}, and editing that selects what to
remove by the text \emph{word targeting}. We propose it as an ideal type: a mechanism with checkable predictions,
silent on any editor's intent. Word targeting reads the sense of a word in its sentence, which a lexicon cannot do, so ``a minority of cells'' is spared. What it reads, though, is the text. Three implications
follow, and each can fail.

First, what is removed is selected by the text. An edit can be aimed at four levels: the \emph{sentence} a name sits
in, the \emph{form} of the name, the \emph{population} named, and the \emph{research}, that is, whether the
project studies the population. Word targeting predicts that removal follows the first three levels more than the
fourth. Because research on a population also shapes how often and where it is named, the prediction can be tested
only as a question of added information: whether practice records predict removal once the text is known. Where era vocabulary and informative names share sentences, edits aimed at the vocabulary will also remove
names.

Second, the count inherits the edits. A category mined from the edited text moves with the text, whoever made the
edits and without requiring that anyone game the category. Where the edited words had been informative, the result is \emph{uncounted science}. It can become a constitutive effect, in the sense of \citet{dahlerlarsen2014},
if the record is then read, and acted on, as a retreat of the research.

Third, word targeting has a boundary. Under \emph{population targeting}, where the pressure is aimed at the population
itself, editing words cannot satisfy it. Distinctions between forms of the population's name should
then stop mattering, and the action should move from the text to the award. The 2003 congressional challenge to NIH grants for
research on sexuality and HIV is the precedent for population targeting, though not for termination: no grant was
rescinded then \citep{kempner2008,kempner2011}.

Table~\ref{tab:tests}, after the Results, lists these implications, two more (that the names were informative, and that practice records changed less than the summaries), and the evidence for each.

\section{Setting and data}
\label{sec:setting}

\subsection{The 2025 reviews}

Executive orders of January 20 and 21, 2025, addressed DEI programs and ``gender ideology''
\citep{eo14151,eo14168,eo14173}. NIH's guidance of March 4, 2025, sorted awards by the role of DEI in the project and
required removal of ``DEI language'' from awards that did not support DEI activities (Introduction). It is reported to
have barred the word ``LatinX'' as not clearly defined \citep{salles2026}. The June 16, 2025, ruling vacated the
directives, and on August 21, 2025, the Supreme Court left the vacatur of the directives in place while staying the part of the ruling that set aside terminations \citep{scotus2025}. NIH continued to
issue delayed continuations through the end of the fiscal year. In December 2025 NIH introduced a text-analysis
screen, reported in June 2026 to check applications against at least 235 terms \citep{kozlov2026}. A September 2026
class action states that the tool ``was created from a database of the words that appeared frequently in the grants
terminated in 2025'' \citep[complaint ¶143]{salles2026}. The 235-term list thus postdates the edits studied here and,
by that account, was built from the words of grants terminated in 2025. We use it only to describe the event.

\subsection{The vocabulary the review would act on}

The share of new NIH awards containing a screen term rose from 6.4\% in FY2015 to 15.8\% in FY2024. The rise
occurred within the same contact principal investigators (+9.7 percentage points; 95\% CI, 8.7–10.6) and in parallel
at NSF (29.2\% in 2015–2019 to 38.4\% in 2023–2024). Population naming rose too, but broadly. Title naming of
caregivers, a comparison population, rose 5.9-fold, against 2.8-fold for Hispanic populations and 1.4-fold for Black
populations. What was specific to the screen was a set of stems, such as \emph{equit-}, \emph{underrepresent-} and
\emph{minorit-}. Their rise within the same investigators and across agencies suggests that they mark a period of writing more than a kind of research. By 2024, population names that
recorded study design sat in the same sentences as this era vocabulary. Those shared sentences are what the 2025
review would act on.

\subsection{Data}

\emph{Awards.} NIH ExPORTER project and abstract files for FY2015–FY2025 provide titles, abstracts, public health
relevance statements, award notice and budget start dates, costs and spending categories. We restrict to
NIH-administered extramural grants with a cost, excluding subprojects and supplements, and keep the latest entry per
core project and fiscal year.

\emph{Papers.} OpenAlex award identifiers (snapshot of June 26, 2026) link core projects to 1.25 million articles
published 2016–2026, with Medical Subject Headings (MeSH), the subject descriptors assigned by the National Library of Medicine (NLM). Because NLM moved all MEDLINE indexing to
automated methods in 2022, we treat MeSH on articles published in 2021 or earlier as curated and do not use later
MeSH as evidence of practice. Publication year approximates the date of indexing.

\emph{Trials.} ExPORTER's clinical studies file links awards to trials. Aggregate Analysis of ClinicalTrials.gov
(AACT) snapshots of January 1, 2025, and September 25, 2026, give status, eligibility criteria, new registrations and
posted baseline counts by race, ethnicity and sex.

\emph{The count.} RCDC categories are text-mined from the title, abstract, public health relevance statement and
specific aims \citep{rcdc}. We use Racial and Ethnic Minority Health Research (REMHR), whose assignment among
unchanged summaries was stable across years. The Health Disparities Research category changed among unchanged
summaries in FY2025, consistent with a change of definition, and is not used as an outcome. REMHR is defined as
``the scientific investigation of distinctive health characteristics and attributes of minority racial/ethnic groups
(including multi-racial individuals)'', and awards that simply stratify data by race or ethnicity are excluded
\citep{grason2024}. To describe where
projects went after losing REMHR, we group five RCDC categories as \emph{equity categories}, a grouping of ours:
Health Disparities Research; Health Disparities and Racial or Ethnic Minority Health Research; Social Determinants of
Health; American Indian or Alaska Native; and Coronaviruses Disparities and At-Risk Populations.

\emph{Terminations.} Grant Witness's list of NIH awards terminated, frozen or reinstated (file of March 23, 2026) gives
status and termination dates.

\subsection{Terms}

\emph{Screen terms} are the 235-term list. For tests of what was deleted or terminated we use the 699 terms in
Appendix~B of the US Senate Commerce Committee minority staff report on NSF grants, published October 9, 2024, before
the edits \citep{senate2024}. We call this the Senate list, and a term or form on it \emph{listed}. \emph{Flagged}
awards, a label of ours, contain a screen term. The \emph{summary} of an award is its title, abstract and public
health relevance statement. \emph{Practice records} are records the summary does not produce: enrollment, curated
MeSH and trial eligibility. Six populations are \emph{signaled}, meaning that the 2025 guidance or orders named them or used vocabulary describing them: Hispanic or Latino; Black or African American; American Indian or Alaska Native; Asian; sexual
and gender minorities (SGM); and low income. Eight \emph{comparison} populations carried no such signal: veterans,
military personnel, health personnel, students, athletes, caregivers, twins and prisoners. Women serve as a benchmark in the enrollment analysis, where sex is reported alongside race
and ethnicity. \emph{Revisions} are changes between the FY2024 and FY2025 summaries. Who made them is not observed. The
\emph{count} is a project's REMHR assignment in a year, and the \emph{record} consists of the category totals NIH publishes.

\section{Methods}
\label{sec:methods}

A continuation pair links a project's latest eligible FY2024 entry to its latest FY2025 non-competing (type~5)
entry. Event analyses require an abstract in both years (37,406 pairs; 5,461 flagged). Mention analyses require any
text in both years (37,790 pairs), and their unit is an award and a population named in its FY2024 text. A name survives
if the FY2025 summary still names the population anywhere. A mention shares a sentence with a listed term if its
sentence also contains a listed term other than the mention. Collateral deletion is the removal of such sentences.

A population is \emph{studied} by a project if pre-2025 practice records show that the research involves it: at least
two curated-era articles with at least 30\% carrying the population's MeSH descriptor, a linked trial whose January
2025 eligibility criteria name it, or, for Black and Hispanic populations, a linked trial that enrolled at least twice
the US population share. The standard captures involvement of the population, not the purpose that REMHR is defined
to capture. ``Studied'', or ``strict practice evidence'', is short for meeting it, and projects below it may still
study the population. Tests of
research use the measured subset of projects with at least three curated-era articles or a linked trial, which covers 22.3\% of revised award--population pairs for
signaled populations.

Human judgments were made by three reviewers, called reviewers 1 to 3 below: two of the authors (reviewers 1 and 3) and a colleague (reviewer 2).
Reviewer 1 designed the study, and reviewers 2 and 3 took no part in the data analysis. All knew the study's aims,
No review packet showed project identifiers or category status, and the blinding of each review is given with it.

Retention of the count is, among projects that strictly study a racial or ethnic population and carried REMHR in FY2024, the share that still carry it in FY2025, by what happened to the summary. Among FY2024 awards active after January 20, 2025 (55,302), an award is recorded as terminated if Grant
Witness gives a termination date and a status of terminated or possibly reinstated. Adjusted differences hold the
count of listed terms, institute and activity fixed. Intervals for proportions are Wilson intervals, with a Newcombe interval for the difference in retention. Other intervals resample projects, or institutions where stated, except a normal approximation for the difference between two form gaps. Definitions, models and sensitivity analyses are in Supplementary Methods~\ref{sm:data}--\ref{sm:inference}.

\section{Results}
\label{sec:results}

\subsection{Four in ten flagged summaries lost every screen term in one year}

\begin{figure}[t]
\centering
\figfile{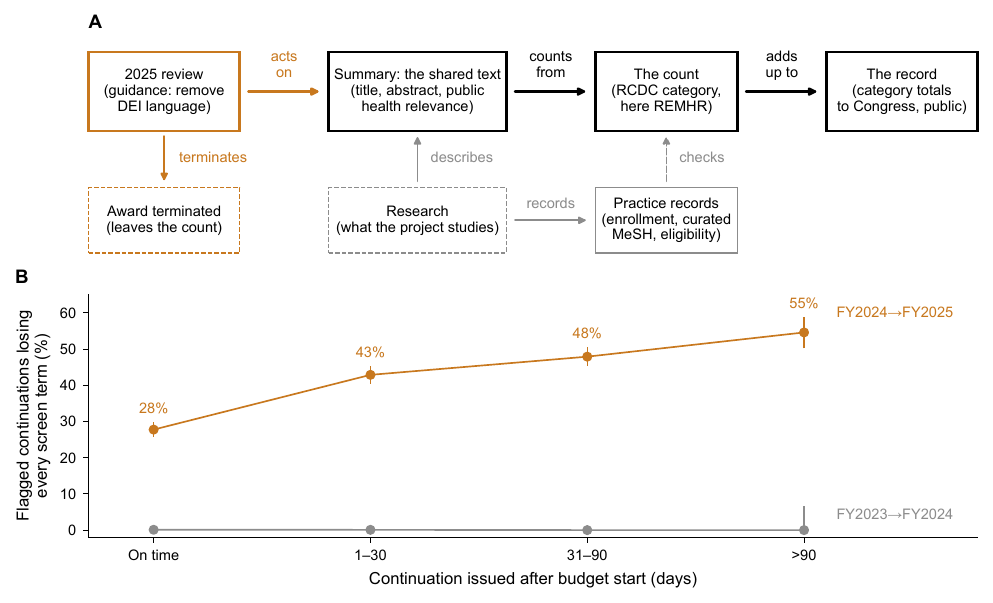}{6cm}
\caption{The review edited the text the count is mined from. (A) The shared text: the review acts on the summary,
the count is mined from it and adds up to the record NIH publishes, and practice records, which the summary does not
produce, record aspects of the research. Terminated awards leave the count. (B) Share of flagged continuations that lost every
screen term, FY2023$\rightarrow$FY2024 and FY2024$\rightarrow$FY2025, by days from budget start to issuance.}
\label{fig:event}
\end{figure}

Of 5,461 flagged FY2024 awards with an FY2025
continuation abstract, 2,169 (39.7\%) contained no screen term, against 4 of 5,242 (0.08\%) a year earlier, when only
62 flagged summaries changed at all (Figure~\ref{fig:event}B). The later the continuation, the more it lost: from
27.8\% of continuations issued on time to 54.6\% of those issued more than 90 days after budget start. Removal was
sensitive to meaning. Summaries whose only hits were incidental uses, such as ``a minority of cells'', lost them in 2.4\% of cases,
against 40.6\% for uses in the listed sense.

Flagged continuations were issued later (65.1\% after June 16, 2025, against 43.9\% of unflagged continuations), and
the gradient by delay held when standardized to the mix of budget-start months (27.3\% to 52.8\%). Among flagged
awards issued before January 20, 2025, 22.8\% of summaries had nonetheless changed, against 0.8\% of unflagged
summaries, which indicates that summaries were revised after issue.

\subsection{Population names recorded whom a study enrolled}

\begin{figure}[t]
\centering
\figfile{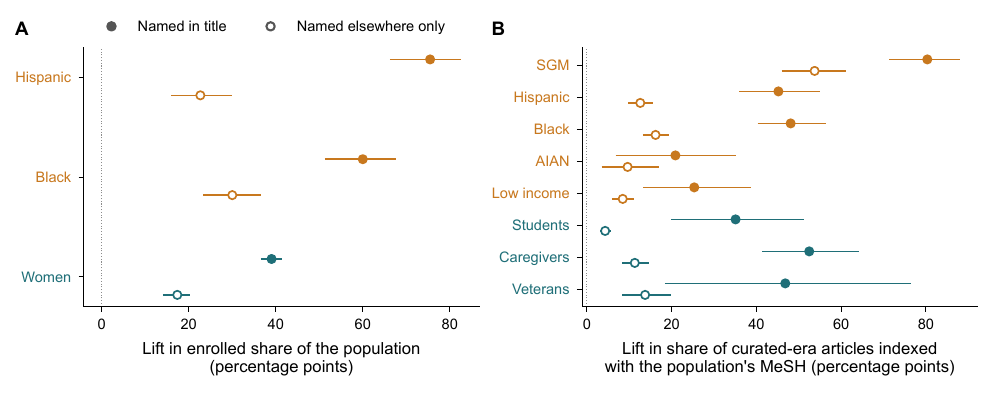}{5cm}
\caption{Population names recorded whom a study enrolled. (A) Lift in the enrolled share of a population in linked
trials, awards naming it against awards not naming it, by placement (filled, title; open, elsewhere only). (B) Lift
in the share of curated-era linked articles indexed with the population's MeSH descriptor. Amber marks signaled populations and teal comparison populations. New awards FY2015–FY2023, with 95\% intervals.}
\label{fig:premise}
\end{figure}

Naming a population in a grant predicted whom its trials enrolled (Figure~\ref{fig:premise}A). Among new awards with
linked trials posting baselines, the enrolled share of Black participants was 60 percentage points higher (95\% CI,
51–68) when the title named Black populations, and 30 points higher (23–37) when only the abstract or public health
relevance statement did, than in awards not naming them. The lifts were 75 and 23 points for Hispanic participants and
39 and 17 points for women. Curated MeSH showed the same gradient (Figure~\ref{fig:premise}B; Black, 48 and 16 points;
caregivers, a comparison population, 52 and 11). Lifts for signaled names were of similar size to those for comparison names. Mentions outside the title were less informative but still predicted enrollment. Awards with linked trials posting baselines were 2\%
of new awards (2,504 of 126,762), and the enrollment results describe that subset.

\subsection{Text features predicted which names survived}
\label{sec:mechanism}

Name survival varied with three of the four levels distinguished in Section~\ref{sec:theory}: sentence, form and
population. At the fourth, research, whether a project studied the population added little to the prediction of
which names survived once text features were known. Because studying a population also makes its name more frequent
and denser, the finding concerns prediction, and it does not separate editing that read the words from editing that
read the research through the words.

\begin{figure}[t]
\centering
\figfile{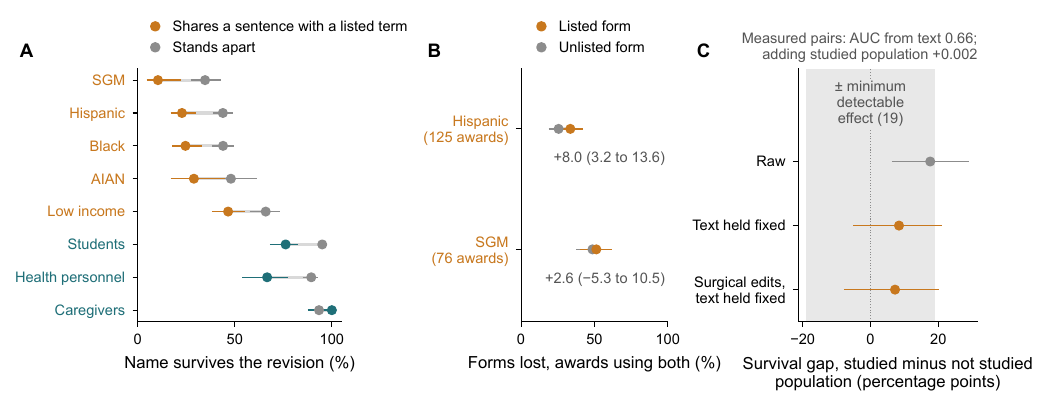}{6cm}
\caption{Text features predicted which names survived. (A) Survival of population names among revised continuations,
when every mention shares a sentence with a listed term and when one stands apart. (B) Within-award loss of listed
and unlisted forms of the same population, with the difference and its 95\% interval. (C) Survival gap for studied
populations, raw, with text held fixed, and for surgical edits, against the minimum detectable effect (shaded).
Intervals are 95\%, Wilson for proportions and bootstrap resampling projects for gaps.}
\label{fig:mechanism}
\end{figure}

\emph{Sentence.} Among revised continuations, a name whose every mention shared a sentence with a listed term
survived in 22.8\% of Hispanic, 24.6\% of Black and 10.4\% of SGM cases, against 43.9\%, 43.9\% and 34.7\% when at
least one mention stood apart (Figure~\ref{fig:mechanism}A). Comparison populations showed a gap at a higher level
(students 76.2\% against 95.0\%; health personnel 66.7\% against 89.4\%). Collateral deletion removed sentences
carrying era vocabulary and any name they held. It fell mainly on workforce-adjacent populations, whose names appear
in sentences about training and the workforce. Pooled, students and health personnel survived in 73.2\% of cases
sharing a sentence with a listed term (95\% CI 66.4–79.1) against 93.4\% when a mention stood apart, while other
comparison populations survived in 89.4\% and 90.8\%.

\emph{Population.} Names of signaled populations were cut out of sentences that otherwise stayed. By an automated sentence
match (Supplementary Methods~\ref{sm:mechanism}), of FY2024 sentences naming a signaled population in revised
continuations, 34.1\% disappeared, 30.4\% remained with the name removed and
35.5\% remained with the name. For comparison populations the shares were 18.0\%, 2.3\% and 79.7\%. Reviewers 3 and 2, blind
to the automated result, checked 100 of these sentences drawn by role and outcome (Cohen's κ = 0.89; reviewer 1
arbitrated 6). Weighted back to all sentences, 47.6\% (95\% CI 39.3–54.7) of signaled sentences remained with the name
removed, against 0.2\% (0.0–0.7) for comparison populations. The automated match understates the contrast, mostly by
counting reworded sentences as gone (Supplementary Table~\ref{tab:s-sentence-check}). In the fixed-effects sample, the mention was lost from 62.5\% of sentences naming signaled populations, and a listed term in the sentence raised the
rate by only 3.0 points. Among comparison populations the rate was 19.4\%, and a listed term raised it by 25.0 points
(award × population fixed effects). Comparison names left mostly with their sentences, and signaled
names on their own.

\emph{Form.} Within awards that used both, listed forms of Hispanic were lost more often than bare ``Hispanic'' or
``Latino'' (Figure~\ref{fig:mechanism}B; 33.6\% against 25.6\%; difference 8.0 points, 95\% CI 3.2–13.6). For SGM, no form difference was detected, and the Hispanic and SGM gaps differ by 5.4 points (−4.1 to 14.8).

\emph{Research.} In the measured subset (332 revised award–population pairs; 110 with a studied population), signaled populations
that the project studied survived more often in raw terms (55.5\% against 37.8\%; +17.6 points, 95\% CI 6.3–29.2).
Studied populations were mentioned more often and more densely, and after
adjustment for mentions, density, listed terms in the sentence, population and institute, the gap was +8.4 points (−5.1 to
21.1; minimum detectable effect 19 points; Figure~\ref{fig:mechanism}C). Corrections for verification and for
misclassification of studied status gave +10.2 to +17.7 points (Supplementary Methods~\ref{sm:mechanism}). The association that remains is estimated imprecisely, and one of moderate size cannot be excluded.
In surgical edits the adjusted gap was +7.2 points. In full rewrites, which change at least half the sentences, the raw gap was +43 points (95\% CI 10–75; 43 pairs, too few to adjust). On all 1,488 revised award–population pairs, a model using only features of the FY2024 text
predicted which signaled names survived with an out-of-fold area under the curve (AUC, where 0.5 is chance) of 0.69 (95\% CI 0.66–0.72). On the 332 measured
pairs its AUC was 0.66, and adding whether the project studied the population changed it by 0.002 (−0.010 to 0.013).

\subsection{Funded projects lost the category when their summaries lost the name}

\begin{figure}[p]
\centering
\figfile{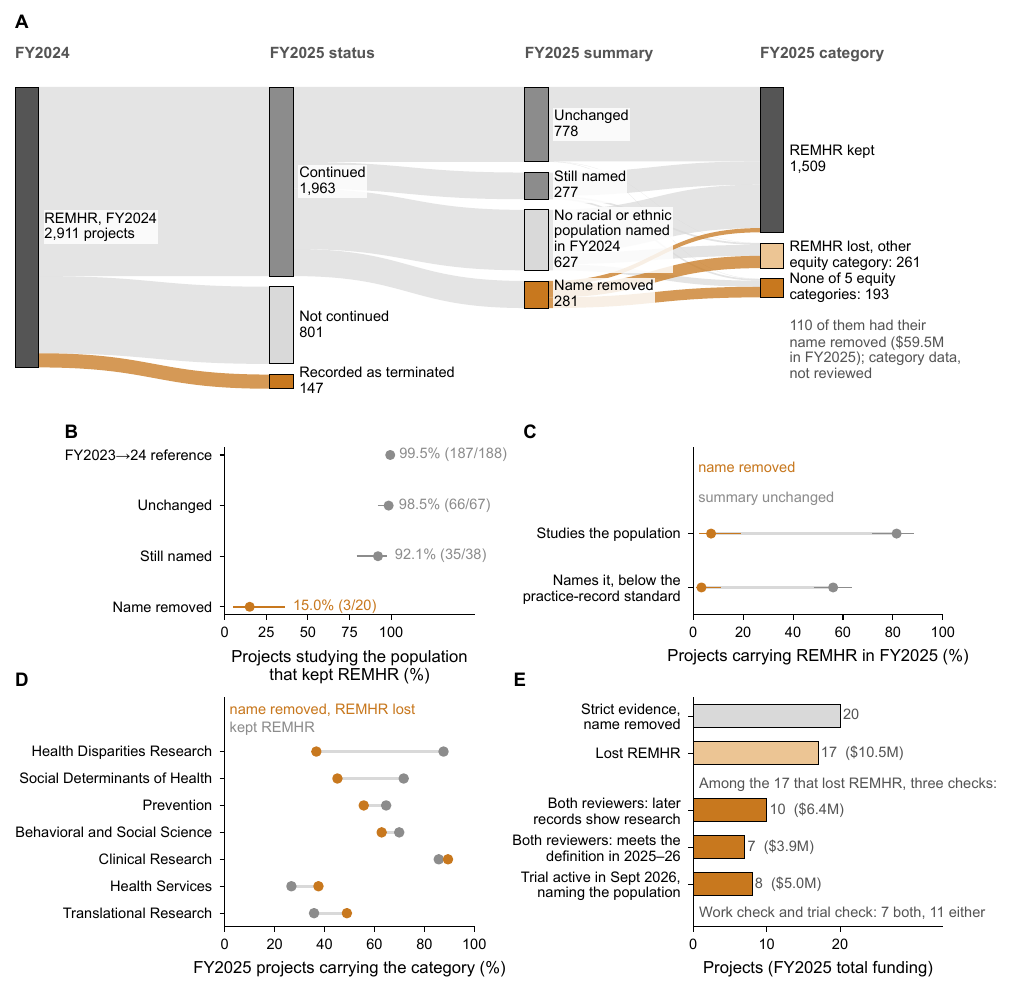}{9cm}
\caption{The projects stayed funded, but the count stopped recording them. (A) FY2024 projects carrying REMHR and active
after January 20, 2025, followed to FY2025: status, the summary, and the FY2025 category. Band widths are projects.
Amber bands are projects whose summaries lost the name. (B) Retention of REMHR among projects that strictly study a racial
or ethnic population, by what happened to the summary, with FY2023$\rightarrow$FY2024 as reference. (C) Share
carrying REMHR in FY2025, summary unchanged (grey) against name removed (amber), for measured projects that study the
population and for those that name it but fall below the practice-record standard. (D) Share of FY2025 projects carrying each category,
projects that kept REMHR against projects that lost it after the name was removed. (E) The 20 continuing projects
with strict practice evidence whose summaries lost the name, and the 17 of them that lost REMHR. Among the 17, three
checks: both reviewers found later public records still showing work on the population (10), both judged the
2025–2026 evidence to show research meeting the category's definition (7), and a linked trial was listed as
recruiting, active or enrolling by invitation in its September 2026 registration record and named the population
(8). The work check and the trial check overlap (7 met both, 11 at least one). Dollars are the projects'
FY2025 total funding, not spending on research on the population, rounded to \$0.1 million. Panel A rests on category
data alone, and panel E on project-level review. Intervals in B and C are 95\% Wilson intervals.}
\label{fig:count}
\end{figure}

The count followed the names (Figure~\ref{fig:count}B). Among projects that strictly study a racial or ethnic population and carried REMHR, 99.5\% kept the category from FY2023 to FY2024 (187 of 188; 95\% CI 97.0–99.9). From FY2024 to FY2025, among 142 continuing projects (including 17 whose FY2024 text named no such population), retention was 98.5\% for unchanged summaries (66 of 67; 92.0–99.7) and 92.1\% for
summaries revised but still naming the population (35 of 38; 79.2–97.3). For summaries that no longer named it,
retention was 15.0\% (3 of 20; 5.2–36.0), 83.5 points below unchanged summaries (95\% CI 61.5–93.3). The 17 projects
that lost the category had \$10.5 million in FY2025 funding, measured as total award costs and not as spending on minority health research.

A blinded review tested whether these projects belong in the category. Each of the 20, with 20 unchanged and 10
still-named controls, was judged against the REMHR definition, which excludes awards that simply stratify data by race
or ethnicity. Packets held the FY2024 summary and linked articles and trial records from before and after 2025, without project identifiers, FY2025 text or category status (Supplementary Methods~\ref{sm:count}). Reviewers 1 and 2 judged every item independently. They agreed on all 50 judgments
of category membership and on 42 of 50 judgments of current work (Cohen's κ = 0.26). The counts below are those on
which both agreed. Both judged that all 20 of the
projects whose summaries lost the name met the definition, as did every control. Of the 17 that lost the category,
both judged that later public records (articles of 2025–2026 or trial records) still showed research on the
population for 10, with \$6.4 million in FY2025 funding. Reviewer 1 alone judged 2 more so, and reviewer 2 alone 3
more. Read from the records alone, 8 of the 17 had a linked trial listed as recruiting, active or enrolling by
invitation in its September 2026 registration record and naming the population in its eligibility criteria. The two
checks overlap: 7 projects met both, and 11 met at least one (Supplementary
Table~\ref{tab:s-audit-cross}).

A second question, asked of the 17 with 13 controls, put the category's own criterion to the 2025–2026 evidence alone: did it show research on the
group's distinctive health characteristics, involvement only, or nothing usable? The reviewers agreed on 20 of 30
items (Cohen's κ = 0.36). Of the 17, both judged 7 to show such research (\$3.9 million in FY2025 funding) and 2
to have no usable evidence, and they disagreed on 8. For 3 of the 7, neither judged the decisive article or trial to
report work done before 2025. Among
the 12 controls that kept the category, 10 showed such research. The stricter criterion narrows the case to 7
projects (Supplementary Table~\ref{tab:s-audit-current}). Reviewers 3 and 2, with reviewer 1 arbitrating, also judged when the work was done (Supplementary
Table~\ref{tab:s-recheck-timing}). For 5 of the 7 (\$2.2 million
in FY2025 funding), the evidence of ongoing work was a trial still listed as recruiting or active in September 2026. For 4 of these 5,
the trial record was actively maintained in 2025–2026 (Supplementary Table~\ref{tab:s-trial-activity}).
Of the other two, one reported work completed before 2025 and one gave no dates. None of the 7 had a 2025–2026 article dating the work to 2025 or later.

A random sample tests how far the result reaches beyond the 20 strict projects. Reviewers 3
and 2 independently reviewed 50 of the 220 other projects that lost REMHR after the name was removed, drawn at random,
mixed with 10 controls that kept it, using packets of the same content. Reviewer 1 arbitrated the 11 items on which
they differed. Before arbitration they agreed on 58 of 60 judgments of FY2024 purpose (Cohen's κ = 0.84) and 52 of
60 of 2025–2026 purpose (κ = 0.77). After it, 42 of the 50 (84\%; 95\% CI 71–92) had an FY2024 purpose meeting the
REMHR definition, which, applied to the 220, is about 185 projects (157–202). The other 8 involved the group only, and for
them the loss of the category is consistent with its definition. Evidence of current work is thinner: 9 of the 50
(18\%; 10–31) showed 2025–2026 research meeting the definition, 14 involvement only, and 27 no usable 2025–2026
evidence. Among the 10 controls, all met the definition in FY2024 and 3 showed such research in 2025–2026, so the low
share reflects how few records these projects leave as well as what happened to them
(Supplementary Table~\ref{tab:s-audit-expand}). Asked when the work was done, the reviewers found that 8 of the 9
reported work done before 2025 and the ninth gave no dates.

Reviewers 3 and 2 also compared the FY2024 and FY2025 summaries of the 70 projects whose summaries lost the name (the 20 strict projects, 3 of which kept REMHR, and the random 50), mixed with the 10 strict projects still naming the population, without project identifiers or
category status. Reviewer 1 arbitrated the 36 of 80 items on which they differed (Supplementary
Table~\ref{tab:s-recheck}). In 60 of the 70 the FY2025 summary no longer referred to the population, in 7 it
did so only through a broader term such as ``underserved'', and in 3 it still named or specifically described it.
The edit recast an aim, exposure, outcome or intervention concerning the population in 49, changed only the wording
in 20 and rewrote the summary in 1 (Cohen's κ = 0.47 before arbitration). Read alone, 63 of the 70 FY2025 summaries
no longer met the REMHR definition, against 2 of the 10 controls (κ = 0.78). The 3 that still met it had unchanged
aims, and in 15 more with unchanged aims the removal of the name alone carried the text out of the definition. The
category read the edited text correctly, and in most cases the edits had changed the stated aims along with the names.

The practice records that define the strict projects were
fixed before 2025, so the category could not have shaped them. The same review may also have changed specific aims or the handling of the category, neither of which is public. The pattern shows that the count tracked the edited text, not that the public summary alone caused the
change. Across all 1,979 continuing projects that carried REMHR in FY2024, features of the text edit alone predicted
loss of the category with an AUC of 0.95 (95\% CI 0.94–0.96). This set includes 16 whose FY2024 entry gave a project
end in 2024, which Figure~\ref{fig:count}A excludes by starting from projects active after January 20, 2025. Among the 364 measured projects the AUC was 0.938 from
text alone and 0.936 with studied status added (difference −0.002; −0.010 to 0.005).

Checked against practice records, the category fell with the name on both sides of the standard
(Figure~\ref{fig:count}C). Among
measured continuing projects that studied a racial or ethnic population, 81.5\% carried the category in FY2025 when the
summary was unchanged and 7.1\% when it no longer named the population. Among measured projects that named such a
population but fell below the practice-record standard, 56.1\% carried it when the summary was unchanged and 3.3\% when
the name was removed. Projects below the standard may still meet the category's definition, so these losses do not
show that the category had counted them in error.

The funding stayed with the same projects, and what they lost was the specific label (Figure~\ref{fig:count}A). Of 2,911
FY2024 REMHR projects active after January 20, 2025, 1,963 continued. Of the 281 continuing projects whose summary no
longer named a racial or ethnic population, 44 kept REMHR (15.7\%, against 773 of 778, or 99.4\%, whose summary was
unchanged), 127 lost it but still held at least one of the five equity categories, and 110, with \$59.5 million in
FY2025 funding, held none of the five. Projects that lost REMHR kept general research categories such
as clinical, behavioral and translational research (Figure~\ref{fig:count}D), which RCDC assigns alongside specific
ones. Among projects that kept REMHR, 87.7\% also
carried Health Disparities Research, and among those that lost it after the name was removed, 36.7\% did. Unchanged summaries are not a randomized comparison: their awards may have
been reviewed less closely, so the contrast describes what happened to the count and not what each project would have
kept without the edit.

In the analysis universe, REMHR total award costs fell by \$553.7 million
from FY2024 to FY2025 (Supplementary Table~\ref{tab:s-decomposition}). Awards absent from FY2025 accounted for
\$576.8 million, of which \$71.7 million belonged to awards recorded as terminated, and awards new to the universe
added \$312.2 million. Continuing projects that lost the category after their summary stopped naming a racial or
ethnic population carried \$125.0 million in FY2024, about a fifth of the net fall.

The project files we use track the published record. Every project carrying REMHR in the FY2024 and FY2025 files
(4,615 and 3,097) is on NIH's published RCDC project list for that year, which holds 3 and 9 more, and retention computed from those lists is identical (Supplementary Table~\ref{tab:s-frozen}). In dollars, REMHR totals from all records
in the files were 83.9\%, 83.6\% and 82.5\% of the totals NIH published for FY2023, FY2024 and FY2025, and totals from
the analysis universe 71.9\% and 68.1\% in FY2024 and FY2025, consistent with the broader coverage of the published
totals. The published total fell from \$2,612 million in FY2024 to \$1,944 million in FY2025. The decomposition above covers the analysis
universe, not the published total (Supplementary Methods~\ref{sm:data}; Supplementary Table~\ref{tab:s-records}).

\subsection{Awards naming sexual and gender minorities were terminated more often, after adjustment for listed terms}

\begin{figure}[t]
\centering
\figfile{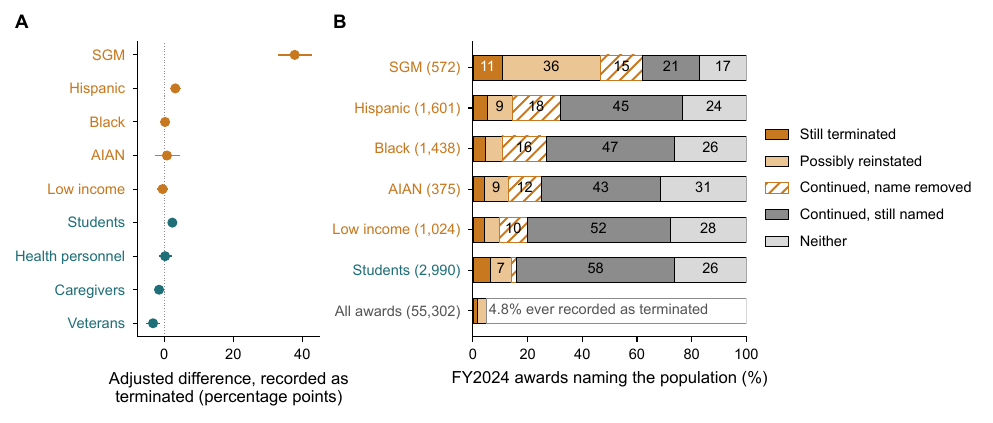}{5cm}
\caption{Awards naming sexual and gender minorities were terminated more often, after adjustment for listed terms. (A) Adjusted difference in the share recorded
as terminated, by population named, holding listed terms, institute and activity fixed, with 95\% intervals resampling
institutions. (B) Outcomes of FY2024 awards naming each population (awards active after January 20, 2025), with all
awards as reference. Numbers in bars are percentages.}
\label{fig:boundary}
\end{figure}

Of FY2024 awards naming sexual and gender minorities and active after January 20, 2025, 46.7\% were recorded as
terminated at some point, against 4.8\% of all awards (Figure~\ref{fig:boundary}B). Of all SGM-naming awards, 10.7\%
were still recorded as terminated and 36.0\% as possibly reinstated (as of March 23, 2026). A further 15.4\% continued with the name
removed, 20.8\% continued still naming the population, and 17.1\% were neither recorded as terminated nor continued.
Holding the count of listed terms, institute and activity fixed,
SGM-naming awards were recorded as terminated 37.8 points more often (95\% CI 33.1–42.7, resampling 2,485
institutions), against +3.2 (1.6–4.9) for Hispanic and +0.2 (−1.2 to 1.8) for Black populations
(Figure~\ref{fig:boundary}A; Supplementary Table~\ref{tab:s-term-adj}). The SGM contrast held for HHS-reported terminations (45.8\% against 3.2\% of all awards), after excluding institutions subject to institution-wide actions (45.0\% against 3.0\%), and among non-training awards (51.4\% against 3.7\%). It also held with
indicators for the 50 most common RCDC research categories (+37.4 points, 32.6–42.1) and with listed terms split into
those inside population names and all others (+31.9, 26.7–37.2). The excess persisted within awards in the RCDC HIV/AIDS category, where
40.8\% of SGM-naming awards were recorded as terminated against 4.0\% of others (+31.8 points, 25.1–38.6;
Supplementary Table~\ref{tab:s-term-topic}). For awards still recorded as terminated, the excess was +7.0 points
(3.9–10.2).

In the edited awards, listed and unlisted SGM forms were lost at similar rates (51.3\% against 48.7\%; difference 2.6
points, −5.3 to 10.5). Similar loss of both forms is consistent with pressure on the population itself, although the SGM
and Hispanic gaps do not differ significantly. Among revised awards using both gender-identity and sexual-orientation forms, gender-identity forms were lost 20.7 points more often
(95\% CI 12.2–30.5; 82 awards). Terminations of awards naming other signaled populations were largely accounted for by
the vocabulary measures, but those of SGM awards were not. We interpret the excess as population targeting.

\subsection{Summaries changed more than the practice records we can follow}

Stated research content changed. Reviewers 1 and 2, blind to year and order, judged that research content stated in the earlier summary was lacking or redefined in the later one in 46.7\% (95\% CI 38.9–54.6) of 150 pairs that lost every screen term
(Krippendorff's α = 0.84; 95\% CI 0.75–0.92). Of these 70 sampled pairs, both reviewers also found content in the later
summary that the earlier one lacked for 66, consistent with redefinition, and 4 lacked content without replacement.
In 68 pairs neither reviewer found content lacking or redefined, and in 12 the reviewers disagreed. Because a study
population counts as research content, removing a population name alone can register as a change in content. The
coding does not separate that case from changes in aims, outcomes or interventions. The paired review of 70 name losses above does, and found aims recast in 49. NIH's notice on renegotiated awards refers to aims and objectives as well as titles and abstracts
\citep{notod26007}. The sample was drawn by the screen terms, so it describes flagged summaries, a different set from the mention analyses
above.

We followed each continuing award and signaled population named in its FY2024 summary to the award's linked trials,
counting a trial only where it could show the change (Supplementary Table~\ref{tab:s-practice}). Of trial links active
in January 2025, 4.5\% (8 of 179) were terminated, suspended or withdrawn by September 2026 when the summary no longer
named the population, against 2.5\% (13 of 529) when the summary was unchanged (difference 2.0 points, −1.4 to 5.8)
and 0.7\% (1 of 141) when it was revised but still named the population. Of trial links whose January 2025
eligibility criteria named the population, 15.6\% (7 of 45) no longer named it when the summary had dropped the name,
against 4.8\% (5 of 105) when the summary was unchanged (difference 10.8 points, −0.5 to 24.7) and 8.5\% (5 of 59) when
it was revised but still named the population. New trials naming the population were registered for 6.4\% of pairs
whose summary dropped the name and 6.2\% of those unchanged. Projects that dropped a name registered more new trials
overall (0.51 against 0.37 per project; difference 0.14, 0.01–0.27). New registrations naming the population barely differed between summaries that dropped the name and unchanged summaries. Trial endings differed by 2.0 points, an interval (−1.4 to 5.8) that does not exclude more than doubling the 2.5\% rate for unchanged summaries. Eligibility criteria for the same population changed more often, on small numbers. Sensitivity analyses are in Supplementary Methods~\ref{sm:robust}.

\begin{table}[p]
\centering
\caption{Implications, rivals, observed estimates and interpretation.}
\label{tab:tests}
\footnotesize
\setlength{\tabcolsep}{3pt}
\begin{tabularx}{\linewidth}{@{}>{\raggedright\arraybackslash}p{1.9cm}>{\raggedright\arraybackslash}p{2.8cm}>{\raggedright\arraybackslash}p{2.3cm}Y>{\raggedright\arraybackslash}p{2.3cm}@{}}
\toprule
Implication & Prediction & Rival & Observed estimate & Interpretation \\
\midrule
Names were informative &
Naming a population predicts its enrolled share and its curated indexing, including when named only outside the title &
Names are ceremonial \citep{meyer1977} &
Enrolled-share lift, title / elsewhere: Black 60 / 30 points; Hispanic 75 / 23; women (sex benchmark) 39 / 17. Curated MeSH lift: Black 48 / 16; caregivers (comparison) 52 / 11 &
Consistent with informative names; enrollment evidence from 2\% of awards \\
\addlinespace
Text predicted name survival &
Survival depends on the sentence, the form and the population named; practice records add little once text is known &
The edits read the research &
Sentence: survival 22.8\% vs 43.9\% (Hispanic), sharing vs not sharing a sentence with a listed term. Form: listed Hispanic forms lost 8.0 points more (3.2–13.6). Population: mention lost from 62.5\% of sentences naming signaled populations, listed term +3.0. Research: +8.4 points (−5.1 to 21.1), MDE 19. Prediction: AUC 0.69 from text on all pairs; on measured pairs 0.66, and adding studied population +0.002 (−0.010 to 0.013) &
Text features predicted survival; practice records added little \\
\addlinespace
The count followed the edits &
Projects that study a population lose the category when their summary loses its name &
RCDC also reads specific aims and keeps continuing research &
Category kept: 98.5\% unchanged, 92.1\% revised and still named, 3 of 20 no longer named; for 10 of the 17 that lost it, both reviewers found later records still showing research on the population (\$6.4 million); 7 showed research meeting the definition in 2025–2026, for 5 only through a trial still listed as active (\$2.2 million); of 70 name losses, aims recast in 49, and 63 FY2025 summaries no longer met the definition; text edit predicts loss, AUC 0.95 &
The count followed the text; the edits reached stated aims \\
\addlinespace
Terminations by population named &
Under population targeting, listed and unlisted forms removed alike; awards terminated beyond what vocabulary predicts &
Exposure to vocabulary, training mechanisms or institution-wide actions &
SGM forms 51.3\% vs 48.7\%; recorded as terminated +37.8 points (33.1–42.7); others between −3.3 and +3.2 &
Read as population targeting \\
\addlinespace
Practice records changed less than summaries &
Summaries change; practice records change less &
Practice re-coupled to the new text \citep{hallett2010} &
Stated content lacking or redefined 46.7\%; eligibility dropped the same population 15.6\% vs 4.8\%, +10.8 points (−0.5 to 24.7); trial endings +2.0 (−1.4 to 5.8) &
Summaries changed; trial endings barely, eligibility more, in a short window \\
\bottomrule
\end{tabularx}

\raggedright\footnotesize Points are percentage points; intervals are 95\%. MDE, minimum detectable effect at 80\%
power. Signaled populations were named in the 2025 guidance or executive orders, or described in their vocabulary;
comparison populations were not. Listed, on the October 2024 Senate list.
\end{table}

\section{Discussion}
\label{sec:discussion}

\subsection*{An edited record}

For continuing projects with strict practice evidence, the FY2025 category recorded what the edited text said.
Table~\ref{tab:tests} sets each implication beside the evidence for it. In
most reviewed losses the edit recast the stated aims as well as the names, and read alone the new summaries no longer
met the definition. The projects were still funded and kept their general categories. A state sees through its
records and acts on what they show \citep{scott1998}, and here the record was edited while the funding continued. A count,
once published, travels without the conditions that produced it \citep{espeland1998}: a reader of the FY2025 table
cannot see that part of it was mined from edited text.

The edits changed the stated research. Of 70 reviewed name losses, 49 had aims concerning the population recast, against 3 of the 10 still-named controls. In
46.7\% of 150 sampled pairs that lost every screen term, both reviewers found research content stated in the earlier summary lacking or redefined in the later one. These comparisons set one account against another and cannot say whether the
research changed or only the account of it. Records the review did not govern answer only part of that question. For
5 of the 7 strict projects judged to show research meeting the definition in 2025–2026, the evidence of ongoing work
is a trial still listed as active, and for 4 of them the trial record was actively maintained. The 2025–2026 articles of the random sample mostly reported work done before
2025. The category was lost whether the edit recast the aims or changed only the wording (47 and 19 of the 67 that lost it).

The result is uncounted science. It differs from two kinds of missing science. \emph{Undone science} is
research that social movements and civil society call for and that institutions leave undone
\citep{frickel2010,hess2016}. \emph{Forbidden knowledge} is research that researchers, gatekeepers and funders
informally judge too sensitive to pursue \citep{kempner2011}. Uncounted science is funded, shows in other records as
work on the population, and is no longer counted. Unlike the other two, it is a property of the count. It becomes constitutive \citep{dahlerlarsen2014}
when the record is read as evidence of what the research is.

\subsection*{What this adds to research evaluation}

Work on the effects of indicators has mostly followed the measured: researchers, departments and journals that learn
what counts and adjust \citep{espeland2007,derijcke2016}. Constitutive effects arise when an indicator comes to
define the thing it was meant to describe \citep{dahlerlarsen2014}. We observe a change in what the official count makes visible. How users read it, and whether decisions or research practice then change, lie outside our data. Our case adds a route in which the pressure acts on the
text the indicator reads and leaves the indicator untouched. Whoever made the edits, the summaries were changed during reviews that made
release of funds conditional on revised text. The case extends Campbell's \citeyearpar{campbell1979} argument to pressure on an indicator's input.

Work on research classification has asked whether a classification is accurate at a point in time. Independent
mappings of research to the same goals disagree widely \citep{armitage2020}. RCDC's prevention category, which is defined
broadly, identified applied prevention research less well than a classifier built for that narrower purpose
\citep{villani2018}, and an audit of its
pediatric category found 6\% of projects misclassified \citep{forrest2026}. These are audits of a cross-section. The
divergence we document is longitudinal: the same project, with the same practice records, is classified differently in
consecutive years, alongside edits to its text. A check of FY2025 against the text the classifier read would find
the category consistent with it. An audit against independent records could find the divergence between category and records, but not when
or how it arose. Users who follow the series, to compare funding with disease burden
\citep{sampat2013} or to track the agency's own portfolio \citep{brown2022,elena2023}, could read the break as
change. Comparing each project's category with records made for other purposes, across versions of its text, is the
design that can see the divergence and when it arose. The case is an argument, from failure, for the plural and contextual
indicators that work on responsible metrics has urged \citep{rafols2019,wilsdon2015}, and for the regular scrutiny of
indicators that the Leiden Manifesto asks of evaluators \citep{hicks2015}. A single count read from a
single text was blind to what had happened to it.

Studies of research governance ask how policy shapes research content \citep{glaser2016}. The 2025 reviews are a case
in which policy acted first on the account of content, and on the funder's record through it. We can follow the
account closely and the content only briefly.

Work on the 2025 events has measured words and terminations. \citet{mehta2025} showed that the words later screened
left NIH text from late 2024. We ask what the deletions selected and what they did to the funder's count, and we
follow the edits through the end of FY2025, when most losses occurred. \citet{rohner2026} found that economists at
federally funded universities changed the content of their papers as well as their words after early-2025 word lists.
Word targeting and substantive change can coexist, and our reviews point the same way for the summaries (aims recast in 49 of 70 reviewed name losses).
\citet{jaramillo2026} found that keyword flags covered about half of terminated predoctoral fellowships, most of which
manual review judged not to be about DEI or specific populations. Their finding is independent evidence that flags and
content diverge. Terminations fell unequally on investigators \citep{liu2025,oliveira2026,fieldingmiller2026}. We add
the population named in the text, and whether the vocabulary or the population predicted termination.

\subsection*{What the edits acted on}

The 2025 guidance was written to separate research from language. The edits read meaning, sparing incidental uses, but
what they removed was predicted by the sentence a name sat in, by whether its form was listed and by which population
it named. Whether the project studied the population added little to the prediction once these were known (AUC change 0.002, −0.010 to 0.013), although the adjusted survival gap does not exclude a moderate association. The paired review shows that the edits usually went beyond the name: in 49 of 70 name losses they recast the stated aims concerning the population. One reading is that the edits could follow a sentence more easily than a study,
and so often failed to tell era vocabulary from the informative names that shared its sentences. An
alternative is that the same process also changed specific aims or internal classification, in which case the
text marks the change without being its only route. The stability of the category among unchanged summaries (99.4\% kept it)
argues against a general change in how RCDC assigned the category, but not against handling specific to reviewed awards. Judged by the guidance's own
distinction between language and research, the edits' errors fell where the summary and the research diverge
\citep{power1997}. Where they removed informative names, the count changed with them.

\subsection*{From one review to the next}

A class action states that the screen NIH introduced in December 2025 was built from a database of words frequent in the grants terminated in 2025 \citep[complaint ¶143]{salles2026}. We have not seen that
database. If the statement is right, the outcomes of one round of review became the word list of the next.
A second route runs through
investigators, who after 2003 reported removing names from later summaries \citep{kempner2008}. If they do so again,
the count could fall further with no change in the research. FY2026 categories are not yet published. In FY2026 records available by September 30, 2026,
3 of 200 projects whose summaries had lost the name named the population again, against 444 of 496 whose summaries
were unchanged (Supplementary Table~\ref{tab:s-names-return}).

\subsection*{Population targeting in 2003 and 2025}

For sexual and gender minorities, listed and unlisted forms were lost at similar rates, and awards were recorded as
terminated far more often than their count of listed terms predicted (37.8 points more often after adjustment, against 3.2 for Hispanic populations). Most of those terminations (36.0 of the 46.7 percentage points) were listed as possibly reinstated by March 23, 2026.
We read this as population targeting: if pressure is aimed at a population, editing words cannot satisfy it, and the
action moves to the award. The same population was the target in 2003. After a
congressional challenge to NIH grants on sexual behavior, a list of more than 250 grants, later claimed by an
advocacy group and believed by investigators to have been compiled by keyword search, reached NIH through committee
staff. NIH reviewed every grant, found them sound, and rescinded none \citep{kempner2008,jnci2004}. Even so, about half
of surveyed investigators reported removing ``red flag'' words from later titles and abstracts, a step they said
program officers often recommended, and 17\% reported dropping studies \citep{kempner2008}. In 2025 the request came
from the agency, awards were terminated as well as edited, and the funder's count inherited the change.

\subsection*{What the record can no longer show}

The problem lies in how the record is built, whatever the direction of the policy. When the words a funder counts are
also the words it polices, its policy and its record of science become one instrument. An administration that asks
for edits to the shared text, whether to remove era vocabulary or to add it, has its own request reported back to it
as a change in the science. In 2025 the names removed were the names the count needed.

The FY2025 series carries an unmarked break. Part of the fall in the minority health total sits in continuing projects whose summaries no longer named the population. In the analysis universe, projects that lost the category after the name was removed
carried about a fifth of the net fall (\$125.0 of \$553.7 million), and that part cannot be recovered from the published totals. Every reading of the series as a trend, whether of funding against
disease burden, of the agency's own portfolio or of priorities under an administration, inherits the break.

We could see the break only because traces survived elsewhere: NIH's published project lists, the FY2024 summaries,
registrations and indexing made for other purposes. The public record itself keeps neither the earlier text nor the
date of an edit. A reader of the record alone cannot tell a change in the
research from a change in its description.

\subsection*{Limitations and future research}

The test of research is imprecise (Section~\ref{sec:mechanism}). The text model's AUC of 0.69 means that the features we measured miss much of what
editors did. The strict cell of the count rests on 20 projects. The practice-record standard captures
involvement of a population, while REMHR is defined by purpose. The blinded reviews support category
membership for those 20 in FY2024, but they rest on packets of public records, and the evidence of ongoing work rests
mostly on registry status, which investigators update on their own schedules. Reviewers agreed only fairly on current work (κ = 0.26 and 0.36) and moderately on how the stated research changed (κ = 0.47). We report agreed counts for the strict cell and the 150 sampled pairs, and arbitrated counts elsewhere. Specific aims, which RCDC also reads,
are not public. The FY2025 files were regenerated in 2026, so edit dates are unknown. Enrollment evidence comes from
the 2\% of awards with trials posting baselines. Grant Witness is a third-party compilation, and termination status
changed with litigation. Above all, the window for practice change is short. Effects on what is proposed, whom
studies enroll, and who is trained take years to appear, and these data cannot show whether the reviews changed them. Practice records followed over a longer window, including new applications and
enrollment, can show whether the reviews changed the research as well as its record.

\section{Conclusion}

NIH's 2025 reviews edited the words of award summaries, and the federal record of science changed with them. We
observe continuing projects that kept their funding while their public summaries lost a population's name and NIH's
annual list dropped the category. In most of the losses we reviewed, the edit recast the stated aims as well as the name, and read alone the new summaries no longer
met the category's definition. For a few, a trial still listed as active is the evidence that work on the population goes on. The funding continued, but the count did not.
Where the target appeared to be the population itself,
awards naming it were terminated far more often. A classification that reads text
is only as valid as the governance of that text. A government that polices research through the text it also counts
can read its own policy back as a fact about science, and so can everyone who relies on its record.

\section*{Data and code availability}

Inputs are public: NIH ExPORTER (files downloaded September 25, 2026); NIH's RCDC project lists and totals (downloaded
September 27, 2026); the NIH RePORTER and ClinicalTrials.gov APIs (queried September 30, 2026); OpenAlex (snapshot of June 26, 2026); AACT (snapshots of January 1, 2025, and
September 25, 2026); Grant Witness (file of March 23, 2026); and the Senate Commerce Committee report of October 9,
2024. The code, with the aggregate results it produces and a script that checks every number printed in the article against
them, and the labels of every human review, with the codebooks, are provided to the editors and reviewers. Award-,
paper-, trial- and person-level files and lists of individual awards are not publicly released.

\section*{Funding}

None.

\section*{Acknowledgments}

We thank Chenxi Xiong (Purdue University), who served as reviewer 2.

\section*{Author contributions}

F.X.: Conceptualization; Formal analysis; Investigation; Writing -- original draft; Writing -- review \& editing. J.Z.: Resources; Data curation; Writing -- review \& editing. H.W.: Conceptualization; Data curation; Formal analysis; Investigation; Methodology; Resources; Software; Validation; Visualization; Writing -- original draft; Writing -- review \& editing.

\clearpage
\setcounter{section}{0}
\setcounter{table}{0}
\setcounter{figure}{0}
\renewcommand{\thesection}{\arabic{section}}
\renewcommand{\thetable}{\arabic{table}}
\renewcommand{\thefigure}{\arabic{figure}}
\renewcommand{\figurename}{Supplementary Figure}
\renewcommand{\tablename}{Supplementary Table}
\makeatletter
\renewcommand\@seccntformat[1]{\@ifundefined{#1@cntformat}{\csname the#1\endcsname\quad}{\csname #1@cntformat\endcsname}}
\def\section@cntformat{Supplementary Methods \thesection.\quad}
\makeatother
\begin{center}
{\Large\bfseries Supplementary Information}\\[4pt]
{\normalsize  Fangfang Xie, Jingwen Zhang and Haining Wang.  Still funded, no longer counted}
\end{center}
\noindent\textbf{Contents.} Supplementary Methods~1--9 give the methods in full, each with the supporting results for
that analysis. Supplementary Figures~\ref{fig:s-sentences}--\ref{fig:s-auc} and Supplementary
Tables~\ref{tab:s-sentence-check}--\ref{tab:s-estimands} follow. Terms are those of the main text. The \emph{summary} is an
award's title, abstract and public health relevance statement. \emph{Practice records} are records the summary does
not produce (trial enrollment, curated MeSH, trial eligibility). A population is \emph{studied} by a project if
pre-2025 practice records meet the standard in Supplementary Methods~\ref{sm:mechanism}. The \emph{count} is a
project's REMHR assignment in a year, and the \emph{record} consists of the category totals NIH publishes.

\noindent\textbf{Where to find each analysis.}

{\scriptsize
\noindent\begin{tabular}{@{}p{2.8cm}p{1.9cm}p{1.7cm}p{3.4cm}p{3.6cm}@{}}
\toprule
Analysis & Main text & Methods & Estimates & Checks and sensitivity \\
\midrule
Event: summaries lost every screen term & Figure~\ref{fig:event} & SM~\ref{sm:data}, \ref{sm:words} & Main text & SM~\ref{sm:robust} \\
Names were informative & Figure~\ref{fig:premise} & SM~\ref{sm:premise} & Main text & SM~\ref{sm:premise}; Supp.\ Table~\ref{tab:s-nulls} \\
Text predicted name survival & Figure~\ref{fig:mechanism} & SM~\ref{sm:mechanism} & Supp.\ Tables~\ref{tab:s-sentences}, \ref{tab:s-forms}, \ref{tab:s-auc}; Supp.\ Figs~\ref{fig:s-sentences}--\ref{fig:s-auc} & Supp.\ Tables~\ref{tab:s-thresholds}, \ref{tab:s-ipw}, \ref{tab:s-tipping}, \ref{tab:s-lca} \\
The count followed the edits & Figure~\ref{fig:count} & SM~\ref{sm:count} & Supp.\ Tables~\ref{tab:s-audit}, \ref{tab:s-audit-cross}, \ref{tab:s-validity} & Supp.\ Tables~\ref{tab:s-records}, \ref{tab:s-frozen}, \ref{tab:s-decomposition} \\
Terminations by population named & Figure~\ref{fig:boundary} & SM~\ref{sm:terminations} & Supp.\ Table~\ref{tab:s-term-adj} & Supp.\ Tables~\ref{tab:s-term-defs}, \ref{tab:s-term-topic} \\
Practice records & Section~\ref{sec:results} & SM~\ref{sm:practice} & Supp.\ Table~\ref{tab:s-practice} & SM~\ref{sm:practice} \\
Estimands, inference, reproducibility & Section~\ref{sec:methods} & SM~\ref{sm:inference} & Supp.\ Table~\ref{tab:s-estimands} & SM~\ref{sm:inference} \\
\bottomrule
\end{tabular}}

{\scriptsize SM, Supplementary Methods; Supp., Supplementary. Figures and Sections without a prefix are in the main text.}

\section{Data, samples and time ordering}\label{sm:data}

\emph{Awards.} NIH ExPORTER project and abstract files for FY2015–FY2025 (downloaded September 25, 2026) give titles,
abstracts, public health relevance statements, award notice and budget start dates, costs and RCDC categories. We
restrict to NIH-administered extramural grants with a cost, excluding subprojects and supplements, and keep the latest
entry per core project and fiscal year. A continuation pair links a project's latest eligible FY2024 entry to its
latest FY2025 non-competing (type~5) entry. Event analyses require an abstract in both years. Mention analyses require
any text in both years, and their unit is an award and a population named in its FY2024 text (an award–population
pair). An award is revised if its summary changed. Mention sentences are sentences of the abstract and public health
relevance statement that name the population.

\emph{Record versions.} The FY2025 project files were regenerated by NIH in 2026, so edit dates are unknown. REMHR
assignments in the files match NIH's published RCDC project lists for REMHR (category 335, downloaded September 27, 2026):
every project carrying REMHR in the FY2024 and FY2025 files is on the list (4,615 and 3,097; the lists hold 3 and 9
more), all 50 reviewed items agree in both years, and retention computed from the lists is identical
(Supplementary Table~\ref{tab:s-frozen}). In dollars, REMHR totals from all records in the files were 82.5–83.9\% of
the totals NIH published on June 23, 2026, and totals from the analysis universe 71.9\% (FY2024) and 68.1\% (FY2025),
consistent with the broader coverage of the published totals. Both fell by about a quarter in FY2025
(Supplementary Table~\ref{tab:s-records}).

\emph{Papers and trials.} OpenAlex award identifiers (snapshot of June 26, 2026) link core projects to articles with
MeSH descriptors. MeSH on articles published in 2021 or earlier is treated as curated, because MEDLINE indexing became
automated in 2022. ExPORTER's clinical studies file links awards to trials, and AACT snapshots of January 1, 2025, and
September 25, 2026, give status, eligibility criteria, new registrations and posted baseline counts.

\emph{Terminations.} Grant Witness's list of NIH awards terminated, frozen or reinstated (file of March 23, 2026) gives
status and termination dates. Its statuses are terminated and possibly reinstated, and it does not confirm
reinstatement. The denominator is FY2024 awards active after January 20, 2025.

\section{Word lists, populations and forms}\label{sm:words}

\emph{Word lists.} Screen terms are the 235-term list reported in June 2026. They describe the event only, because the
list was reportedly built from the words of terminated grants. Inference uses the Senate list, the 699 terms in
Appendix~B of the Senate Commerce Committee minority staff report of October 9, 2024. A mention shares a sentence with
a listed term if its sentence also contains a listed term other than the mention.

\emph{Populations.} Six signaled populations (Hispanic or Latino; Black or African American; American Indian or Alaska
Native; Asian; sexual and gender minorities; low income) and eight comparison populations (veterans, military
personnel, health personnel, students, athletes, caregivers, twins, prisoners) are defined by name patterns and MeSH
descriptors. For SGM, whose current MeSH descriptor was introduced in 2017, we include the descriptors NLM lists as its
previous indexing. Comparison populations have MeSH descriptors that are not check tags. Zebrafish, \emph{Drosophila}
and mice serve as mechanical checks.

\emph{Forms.} For Hispanic and SGM populations we compare forms on the Senate list (for example ``Hispanic
community'', ``LGBTQ'', ``transgender'') with forms not on it (bare ``Hispanic'' or ``Latino'', ``gay'', ``men who
have sex with men'') within awards that used both.

\section{Population names and practice records}\label{sm:premise}

For new awards of FY2015–FY2023 with linked trials that started within three years and posted baselines, the lift is
the difference in the enrolled share of a population between awards naming it and awards not naming it, by placement
(title, or only elsewhere). The analogous MeSH lift uses curated-era articles. Excluding trials whose eligibility named
the population, the enrollment lifts for Black participants were 47 and 23 points. A cohort test of whether naming
became less informative over 2015–2023 was null (Supplementary Table~\ref{tab:s-nulls}).

\section{What the edits removed}\label{sm:mechanism}

\emph{Sentence fate.} Each FY2024 sentence naming a population in a revised continuation is matched to its most similar
FY2025 sentence by token Jaccard similarity. The sentence is gone if no match reaches 0.5. Otherwise it is kept, and
the name is kept or removed according to the matched sentence (Supplementary Table~\ref{tab:s-sentences}). With
thresholds of 0.4 and 0.6, signaled sentences kept with the name removed were 35.5\% and 24.3\%, against 3.1\% and
1.7\% for comparison populations (Supplementary Table~\ref{tab:s-thresholds}). The contrast holds at each threshold. Whole
sentences holding a listed term were more often deleted. Names of signaled populations were also cut from sentences
that otherwise stayed, which almost never happened to comparison populations.

\emph{Human check of the sentence match.} One hundred FY2024 sentences were drawn by population role and automated
outcome (signaled, 20 per outcome; comparison, 15 gone, 10 kept with the name removed and 15 kept with the name; at most one per award and stratum). Each packet gave the population tracked, the FY2024 sentence and the
FY2025 summary as numbered sentences, without identifiers, role or automated outcome. Reviewers 3 and 2 named the
FY2025 counterpart and judged the sentence gone, kept with the name removed, kept with only a broader term, or kept
with the name in any wording. They agreed on the fate of 94 sentences (Cohen's κ = 0.89) and on the counterpart of 87,
and reviewer 1 arbitrated the other 6. A broader term counts as name removed (1 sentence). Shares are weighted back to
the 4,033 signaled and 4,277 comparison sentences by stratum, with intervals resampling items within strata (2,000
draws; Supplementary Table~\ref{tab:s-sentence-check}).

\begin{figure}[htbp]
\centering
\includegraphics[width=0.55\linewidth]{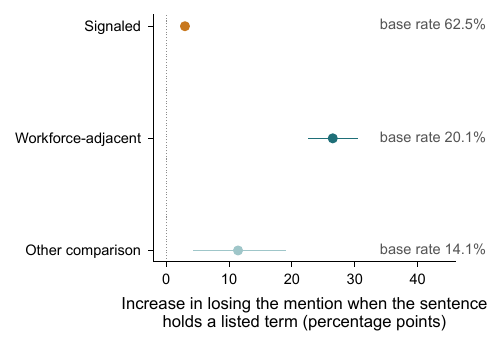}
\caption{Increase in losing the mention when the sentence holds a listed term, within award × population. Comparison
populations are split into workforce-adjacent (students, health personnel) and other. Bootstrap 95\% intervals resample
awards (none shown for signaled populations). Supplementary Methods~\ref{sm:mechanism}.}
\label{fig:s-sentences}
\end{figure}

\begin{figure}[htbp]
\centering
\includegraphics[width=0.55\linewidth]{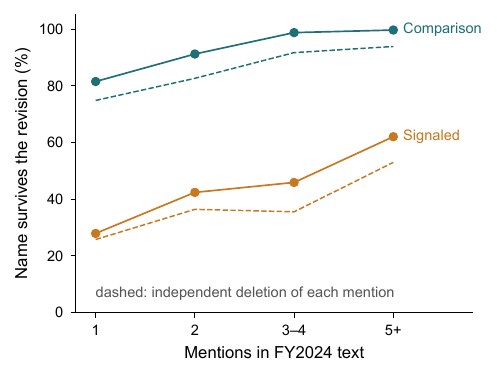}
\caption{Survival of signaled and comparison population names by number of mentions in the FY2024 text, observed
(solid) and predicted under independent deletion of each mention (dashed). Supplementary Methods~\ref{sm:mechanism}.}
\label{fig:s-dose}
\end{figure}

\emph{Sentence model and dose.} A linear probability model of losing the mention, with award × population fixed
effects, includes whether the sentence holds a listed term, whether it states aims or methods, and length. The
increment was 26.5 points (22.6–30.5) for workforce-adjacent populations and 11.4 points (4.3–19.0) for other
comparison populations (Supplementary Figure~\ref{fig:s-sentences}). Survival rose with the number of mentions
(Supplementary Figure~\ref{fig:s-dose}). For signaled populations it went from 28\% with one mention to 62\% with five
or more, and for comparison populations from 82\% to 99.7\%. Awards with five or more mentions kept at least one more
often than independent deletion of each mention predicts (62.0\% against 52.9\%).

\emph{Forms.} The Hispanic and SGM form gaps differ by 5.4 points (−4.1 to 14.8;
Supplementary Table~\ref{tab:s-forms}).

\emph{Practice-record standard.} A population is studied by a project if pre-2025 practice records show that the
research involves it: at least two curated-era articles with at least 30\% carrying the population's MeSH descriptor,
a linked trial whose January 2025 eligibility criteria name it, or, for Black and Hispanic populations, a linked trial
that enrolled at least twice the US population share. The measured subset contains projects with at least three
curated-era articles or a linked trial. REMHR is defined as ``the scientific investigation of distinctive health
characteristics and attributes of minority racial/ethnic groups (including multi-racial individuals)'', and awards
that simply stratify data by race or ethnicity are excluded \citep{grason2024}. The standard therefore records
involvement of a population, not the purpose the category is defined to capture.

\begin{figure}[htbp]
\centering
\includegraphics[width=0.55\linewidth]{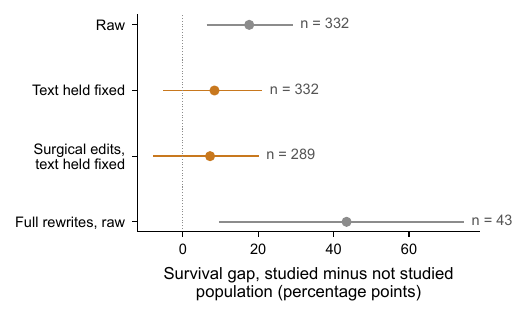}
\caption{Survival gap, studied minus not studied population, raw and with text held fixed, overall, for surgical
edits and, raw only, for full rewrites, with 95\% intervals resampling projects. Supplementary Methods~\ref{sm:mechanism}.}
\label{fig:s-research}
\end{figure}

\emph{Research role.} We report raw and text-adjusted gaps (mentions, density, listed terms in the sentence, population, institute)
and the minimum detectable effect at 80\% power. The text-adjusted gap is an association between practice evidence and
survival with text features held fixed. Because studying a population may itself make its name more frequent and
denser in the text, the adjustment changes the question. 
Weighting measured pairs by the inverse probability of having a practice record gave +10.2 points (−5.4 to 23.9)
(Supplementary Table~\ref{tab:s-ipw}). Because studied-population status is itself measured with error, the estimate is
attenuated. With sensitivity 0.7 and specificity 0.95 for that status, the corrected gap is +11.7 points (upper 95\% limit +29.3), and with the values implied by a latent class model of the three record types it is +17.7 (upper 95\% limit +44.5;
Supplementary Tables~\ref{tab:s-tipping} and \ref{tab:s-lca}). In full rewrites, revisions changing at least half the
sentences, studied populations survived more often (raw +43 points; 95\% CI 10–75; 43 pairs), and the subset is too small to hold text features fixed. In surgical edits the adjusted gap was +7.2 points (Supplementary Figure~\ref{fig:s-research}).

\begin{figure}[htbp]
\centering
\includegraphics[width=0.55\linewidth]{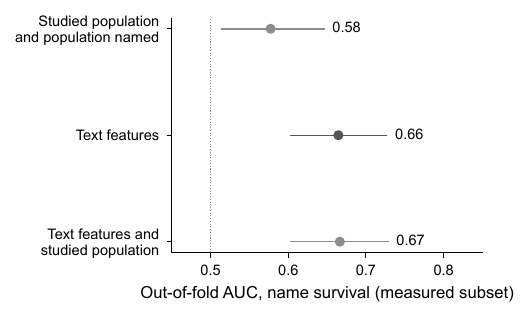}
\caption{Out-of-fold AUC for name survival in the measured subset, by model, with 95\% intervals resampling projects.
Supplementary Methods~\ref{sm:mechanism}.}
\label{fig:s-auc}
\end{figure}

\emph{Prediction.} Out-of-fold AUCs use five folds grouped by project (Supplementary Table~\ref{tab:s-auc}; Supplementary
Figure~\ref{fig:s-auc}). The text model was well calibrated (calibration-in-the-large 0.003; slope 0.85).
The adjusted gap is the more direct test of added value.

\section{The count}\label{sm:count}

\emph{Retention.} Retention is the share of projects that strictly study a racial or ethnic population and carried
REMHR in FY2024 that still carry it in FY2025, by what happened to the summary. Among strict-evidence REMHR carriers whose FY2024 text named no racial or ethnic population (17 projects), retention was 58.8\%.

\emph{Decomposition.} Supplementary Table~\ref{tab:s-decomposition} splits the FY2024–FY2025 change in REMHR total
award costs in the analysis universe into components that sum to it.

\emph{Blinded review of category membership.} The 20 continuing projects that strictly study a racial or ethnic
population, carried REMHR in FY2024 and whose summaries lost the name were reviewed with 20 randomly drawn projects
whose summaries were unchanged and 10 whose summaries were revised but still named the population.
Each packet held the FY2024 summary, up to ten linked articles published before 2025 and up to ten published
2025–2026 (titles and abstracts from PubMed; the ten most recent in each window when more were linked), and linked trials with status, eligibility and enrolled shares in the
January 2025 and September 2026 AACT snapshots. Packets carried no project, application, article or trial
identifiers, no FY2025 text and no category status, and items were shuffled across groups. The codebook asked two questions.
The first was whether the research investigates distinctive health characteristics of a minority racial or ethnic
group so as to meet the REMHR definition (A), involves the group only through enrollment, sampling or stratification
(B), or leaves the question undecided (C). The second was whether 2025–2026 evidence shows work on the group (verified, historical only,
unknown). The first question was answered from the FY2024 summary and evidence dated before 2025, the second from
evidence dated 2025--2026. Reviewers 1 and 2 labeled every packet independently against the codebook, with one label per question for each item. They agreed on the first question for all 50 items, every one judged
to meet the definition, so no chance-corrected statistic is defined. On the second they agreed for 42 of 50 (Cohen's
κ = 0.26; Krippendorff's α = 0.27, 95\% CI −0.09 to 0.64, 2,000 item bootstrap draws). Of the 8 disagreements, 4 were
unknown for reviewer 1 and verified for reviewer 2, and 4 the reverse (3 unknown, 1 historical only). The
main text reports the counts on which both agreed. The labels of both reviewers are in Supplementary
Table~\ref{tab:s-audit}.

\emph{Current eligibility.} A current-eligibility question was put to the 17 projects that lost REMHR after the name was removed,
mixed with 13 controls (the 3 that kept REMHR after the name was removed, and 5 each drawn at random from the unchanged
and still-named groups), presented in a new random order. Using only evidence dated 2025--2026, each
reviewer judged whether it showed research investigating distinctive health characteristics of the group (E),
involvement only (I), or no usable evidence (N). Each also recorded the evidence used and whether the decisive article
appeared to report work done before 2025. A trial still listed as not yet recruiting counted as no evidence. The reviewers agreed
on 20 of 30 items (Cohen's κ = 0.36), on the evidence used for 19 and on earlier work for 26 (Supplementary
Table~\ref{tab:s-audit-current}). Every disagreement ran the same way: reviewer 2 judged E where reviewer 1
judged I (4 items), or E or I where reviewer 1 judged N (6). Of the 17, reviewer 1 judged 7 E, 3 I and 7 N, and
reviewer 2 11 E, 4 I and 2 N. Both judged 7 to show research meeting the definition (\$3.9 million in FY2025
funding; 2 by an article, 2 by a trial, 1 by both, and 2 on which the reviewers cited different evidence), and 2 to have
no usable evidence. For 3 of the 7, neither reviewer judged the decisive article or trial to report earlier work. Among the 12 controls that kept REMHR, 10 were judged E.

\emph{A random sample beyond the strict cell.} Of the 237 continuing REMHR carriers that lost the category after the
name was removed, 220 lie outside the strict cell. Fifty were drawn at random, with 5 that kept REMHR
after the name was removed and 5 with unchanged summaries that kept it, all outside the strict cell, presented in a new random order. Packets had the same content as above. Reviewers 3 and 2 answered, independently, the
purpose question for FY2024 (A, B, C) and the current-purpose question for 2025--2026 (E, I, N), with the evidence
used, whether it reported earlier work, the groups concerned and a written reason for each item. Reviewer 1
arbitrated the 11 items on which any answer differed, recording a reason for each. Before arbitration they agreed on
58 of 60 purpose labels (κ = 0.84) and 52 of 60 current-purpose labels (κ = 0.77). After arbitration, 42 of the 50
sampled losses met the definition in FY2024 (84\%; 95\% CI 71–92), or about 185 of the 220 (157–202), and 8 involved
the group only. None was undecided. For 2025--2026, 9 showed research meeting the definition, 14 involvement only
and 27 no usable evidence. All 10 controls met the definition in FY2024 and 3 showed such research in 2025--2026
(Supplementary Table~\ref{tab:s-audit-expand}).

\emph{Paired review of the two summaries.} The 20 strict projects whose summaries lost the name, the random 50 and
the 10 strict projects still naming the population were reviewed in a new random order. Each packet listed the groups detected automatically in FY2024, the FY2024 sentences naming a racial
or ethnic group, and both summaries in full, with no identifiers or category status. Reviewers 3 and 2 judged
independently whether the detected groups were named as populations; how the FY2025 summary referred to the group
(named, specific wording, only a broader term, none); whether the research concerning the group was described with
the same aims, exposures, outcomes and interventions, recast, or rewritten beyond comparison; and whether the FY2025
summary on its own met the REMHR definition. Reviewer 1 arbitrated the 36 items on which any answer differed. The
automated detection was correct for 63 of the 70 name losses, partly correct for 6 (a group appearing only in
``non-Hispanic white'') and wrong for 1 (Supplementary Table~\ref{tab:s-recheck}).

\emph{Timing of 2025--2026 evidence.} The 37 items that at least one reviewer judged to show 2025--2026 research
meeting the definition, in the current-eligibility review or the random sample, were reviewed in a new random order. Packets held the FY2024 summary, titles of earlier articles, full abstracts of
2025--2026 articles and linked trials with registration date, status in both snapshots, enrollment and eligibility.
Award numbers quoted in abstracts were masked. Reviewers 3 and 2 judged which 2025--2026 evidence qualified and when
the qualifying work was done: dated by an article to 2025 or later, shown only by a trial listed as recruiting,
enrolling or active in September 2026, completed before 2025, or undated. Reviewer 1 arbitrated 13 items
(Supplementary Table~\ref{tab:s-recheck-timing}). For the projects whose evidence of ongoing work was a trial listed as active, the qualifying trials' current
ClinicalTrials.gov records gave the last update posted, the date the status was last verified, the current status and
the start date (Supplementary Table~\ref{tab:s-trial-activity}).

\emph{Synthetic illustrations of the codebook.} Item-level packets are not released, because a quoted summary would
identify its award. The four cases below are synthetic: they were written for this Supplement, describe no award, and
show how the codebook was applied to the kinds of evidence the packets contained. Their topics and populations were
chosen to differ from every reviewed item.
\begin{enumerate}
\item \emph{Meets the definition, with current work verified (A, V).} The FY2024 summary aims to test a school-linked program to prevent dental
caries among Hispanic children, and to examine whether family language use moderates its effect. A linked trial is
recruiting in September 2026, and its eligibility criteria name Hispanic or Latino children.
\item \emph{Involvement only (B).} The summary describes a cohort study of sleep and blood pressure that enrolls a
diverse sample and will report results stratified by race and ethnicity, and no aim concerns a particular group. The
REMHR definition excludes awards that simply stratify by race or ethnicity, so the item is B. The case marks the boundary the reviewers applied.
\item \emph{Meets the definition, with current work unknown (A, U).} The summary states an aim about hepatitis B screening
among Asian American adults. The linked trial is listed as not yet recruiting in both the January 2025 and the
September 2026 snapshots, and no linked article appears in 2025--2026. A registered plan is evidence of intent, not
of work in progress.
\item \emph{A case on which reviewers could differ (A; V or U).} The summary concerns kidney function among American Indian
adults with diabetes. A 2025 article reports results from the parent cohort without naming the group, and the only
active linked trial names a different population in its eligibility criteria. One reading
takes the trial and article as continued work by the project, whereas the other finds no record that the named group is still
studied. Several disagreements on current work were of this kind.
\end{enumerate}

\emph{Continued work among projects that lost the category.} Supplementary Table~\ref{tab:s-audit-cross} crosses the
reviewers' judgments of 2025–2026 work, and evidence read directly from the records, with FY2025 category status for
the 20 projects whose summaries lost the name. Of the 17 that lost REMHR, both reviewers verified 2025–2026 work for 10
(\$6.4 million in FY2025 funding), reviewer 1 for 12 (\$8.1 million) and reviewer 2 for 13. Verified
means that later public records still show research on the population, not that each project's scope was unchanged.
By the records alone, 8 of the 17 had a linked trial listed as recruiting, active or enrolling by invitation in its
September 2026 registration record and naming the population in its eligibility criteria (\$5.0 million), 3 had only
a 2025–2026 article naming it, and 6 had neither. A trial still listed as not yet recruiting does not count. The
reviewers' check and the trial check overlap: 7 projects met both (\$4.6 million), 3 only the
reviewers' check, 1 only the trial check, and 6 neither (\$3.6 million). Costs are rounded to \$0.1 million and suppressed for cells of fewer than 3 projects.

\emph{Agreement with the standard.} Treating REMHR as an index test and the practice-record standard as an imperfect
reference, Supplementary Table~\ref{tab:s-validity} gives sensitivity, specificity and positive predictive value in
FY2024 and, in FY2025, by the fate of the summary. 

\emph{Observed counts.} Of the 20 strict-evidence projects whose summaries lost the name, 17 lost REMHR, with
\$10.5 million in FY2025 funding (total award costs, not spending on minority health research). Of all 281 REMHR
carriers whose summaries lost the name, 237 lost it. These counts include any category losses
that would have occurred without an edit. 

\section{Terminations}\label{sm:terminations}

An award is recorded as terminated if Grant Witness gives a termination date and a status of terminated or possibly
reinstated. We also use HHS-reported terminations, exclusion of institutions subject to institution-wide actions and
non-training awards (Supplementary Table~\ref{tab:s-term-defs}). Adjusted differences come from linear probability
models with institute and activity effects and $\log(1 + {}$count of listed terms$)$, with 2,000 bootstrap draws resampling
institutions (Supplementary Table~\ref{tab:s-term-adj}). The excess for student-naming awards reflects
training awards, which are 73\% of student-naming awards. Among non-training awards the student rate was 6.7\%. Adding indicators for the 50 most common RCDC categories of
the FY2024 award, splitting the count of listed terms into terms inside population names and all other terms, or
restricting to awards inside or outside the RCDC HIV/AIDS category left the SGM excess between +31.8 and +41.2 points
(Supplementary Table~\ref{tab:s-term-topic}).

\section{Practice records and stated research}\label{sm:practice}

Each continuing award and signaled population named in its FY2024 summary forms a pair, grouped by what happened to
the name. We follow the award's linked trials between the AACT snapshots and count a trial only where it could show
the change: an ending among trials active in January 2025, and a change in eligibility among trials whose January 2025
criteria named the same population (Supplementary Table~\ref{tab:s-practice}). Reviewers 1 and 2 coded a random sample of 150
FY2024$\rightarrow$FY2025 pairs that lost every screen term. They saw each pair as two texts in random order without years and judged whether one text lacked or redefined
research content (an aim, population, outcome, exposure or intervention) stated in the other. They practiced on 20 separately drawn pairs, not among the 150. On whether the later text lacked or redefined content stated in the earlier one, both answered yes for
70 pairs, only reviewer 2 for 2, only reviewer 1 for 10, and neither for 68 (agreement 92\%). Of the 70 pairs
both judged to lack content, both also judged 66 to state content the other lacked, consistent with
redefinition. The other 4 lacked content without replacement.  The codebook and instructions are in the code package, and the labels are in the review workbook, both provided to the editors and reviewers.

\section{Sensitivity analyses}\label{sm:robust}

With the screen terms in place of the Senate list, the gap between names sharing a sentence with a listed term and
names standing apart was larger (Hispanic 17\% against 42\% survival), as expected of a list built from the words of grants terminated in 2025. With the Senate list the gap was about 20 points. A narrow 26-term race, ethnicity and health equity list
gave the same event contrast (50.5\% of flagged continuations losing every term, against 0.18\%). Enrollment and MeSH
lifts were similar with ExPORTER publication links in place of OpenAlex award links. Termination contrasts held under
every definition. Adjusting for screen terms in place of listed terms left the SGM excess large (+32.3 points) and removed the small Hispanic excess (−0.4
points). 

\section{Estimands, inference and reproducibility}\label{sm:inference}

Supplementary Table~\ref{tab:s-estimands} states the population, exposure, outcome and contrast of each analysis.
Intervals for proportions are Wilson intervals. Bootstrap intervals use 2,000 draws unless stated,
resampling projects or, for terminations, institutions.

Code, aggregate results and a script that checks every number printed in the article against the
result files are in the code package provided to the editors and reviewers, with the review protocols, codebooks and packet builders. The labels of every human review are in the accompanying review workbook. Award-, paper-, trial- and person-level files and lists of individual awards are not publicly released.

\clearpage
\section*{Supplementary Figures}

\clearpage
\section*{Supplementary Tables}

\newcommand{\TabEstimands}{%
\begin{table}[htbp]
\centering
\caption{Estimands of the five analyses.}
\label{tab:s-estimands}
\footnotesize
\begin{tabular}{@{}p{2.0cm}p{3.8cm}p{2.7cm}p{2.4cm}p{2.9cm}@{}}
\toprule
Analysis & Population (denominator) & Exposure & Outcome & Contrast \\
\midrule
Names were informative & New awards FY2015–FY2023 with linked trials posting baselines (2,504) & Population named (title; elsewhere only) & Enrolled share of the population & Difference in means \\
Text predicted name survival & Revised FY2024→FY2025 continuations naming a signaled population (1,488 pairs; 332 measured) & Studied population (practice records) & Name still present & Risk difference with text held fixed (text-adjusted association); AUC \\
The count followed the edits & Continuing projects that strictly study a racial or ethnic population and carried REMHR in FY2024 (142) & Fate of the summary & REMHR in FY2025 & Difference in retention \\
Terminations by population named & FY2024 awards active after January 20, 2025 (55,302) & Population named & Ever recorded as terminated & Adjusted risk difference; crude risk ratio \\
Practice records & Trials linked to continuing awards naming a signaled population in FY2024 (656 projects) & Fate of the name (removed; revised, still named; unchanged) & Trial ended; eligibility; new registrations & Difference in proportions \\
\bottomrule
\end{tabular}
\end{table}}

\newcommand{\TabAuditExpand}{%
\begin{table}[htbp]
\centering
\caption{Blinded review of a random sample of REMHR losses beyond the strict cell, after arbitration.}
\label{tab:s-audit-expand}
\footnotesize
\begin{tabular}{@{}p{3.0cm}rrrrrr@{}}
\toprule
Group & Items & A & B & E & I & N \\
\midrule
Name removed, REMHR lost (sample) & 50 & 42 (71 to 92) & 8 (8 to 29) & 9 (10 to 31) & 14 (17 to 42) & 27 (40 to 67) \\
Name removed, REMHR kept & 5 & 5 (57 to 100) & 0 (0 to 43) & 2 (12 to 77) & 0 (0 to 43) & 3 (23 to 88) \\
Unchanged, REMHR kept & 5 & 5 (57 to 100) & 0 (0 to 43) & 1 (4 to 62) & 2 (12 to 77) & 2 (12 to 77) \\
\bottomrule
\end{tabular}

\raggedright\footnotesize Counts with 95\% Wilson intervals (\%). A, FY2024 purpose meets the REMHR definition; B, involvement only; E, I, N as in Supplementary Table~\ref{tab:s-audit-current}. Reviewers 3 and 2 labeled independently; reviewer 1 arbitrated the 11 items on which they differed. Before arbitration they agreed on 58 of 60 purpose labels (κ = 0.84) and 52 of 60 current-purpose labels (κ = 0.77).
\end{table}}

\newcommand{\TabAuditCurrent}{%
\begin{table}[htbp]
\centering
\caption{Current eligibility in 2025–2026: items given the same label by both reviewers.}
\label{tab:s-audit-current}
\footnotesize
\begin{tabular}{@{}lrrrrrrr@{}}
\toprule
Group & Items & E & I & N & E cost & I cost & N cost \\
\midrule
Name removed, REMHR lost & 17 & 7 & 0 & 2 & \$3.9M & — & s \\
Name removed, REMHR kept & 3 & 3 & 0 & 0 & \$1.8M & — & — \\
Still named, REMHR kept & 4 & 4 & 0 & 0 & \$1.9M & — & — \\
Still named, REMHR lost & 1 & 0 & 0 & 1 & — & — & s \\
Unchanged, REMHR kept & 5 & 3 & 0 & 0 & \$1.6M & — & — \\
\bottomrule
\end{tabular}

\raggedright\footnotesize E, 2025–2026 evidence shows research on the group's distinctive health characteristics; I, involvement only; N, no usable 2025–2026 evidence. The reviewers agreed on 20 of 30 items (Cohen's κ = 0.36); counts are items given the same label by both. Costs are FY2025 totals rounded to \$0.1M; s, suppressed for cells of fewer than 3 projects.
\end{table}}

\newcommand{\TabAudit}{%
\begin{table}[htbp]
\centering
\caption{Blinded review of REMHR membership by group and reviewer.}
\label{tab:s-audit}
\footnotesize
\begin{tabular}{@{}lrrrrrrrr@{}}
\toprule
Group & Items & A & B & C & V & H & U & REMHR kept \\
\midrule
Name removed & 20 & 20 & 0 & 0 & 15 & 1 & 4 & 3/20 \\
Revised, still named & 10 & 10 & 0 & 0 & 10 & 0 & 0 & 9/10 \\
Unchanged & 20 & 20 & 0 & 0 & 19 & 0 & 1 & 20/20 \\
Name removed, reviewer 2 & 20 & 20 & 0 & 0 & 16 & 2 & 2 &  \\
Revised, still named, reviewer 2 & 10 & 10 & 0 & 0 & 9 & 0 & 1 &  \\
Unchanged, reviewer 2 & 20 & 20 & 0 & 0 & 19 & 0 & 1 &  \\
\bottomrule
\end{tabular}

\raggedright\footnotesize A, meets the REMHR definition; B, involvement only; C, undecided. V, later public records show work on the population (Supplementary Methods~\ref{sm:count}); H, historical only; U, unknown. First three rows, reviewer 1; last three, reviewer 2; both blind to group and category status. REMHR kept is among items judged A (Supplementary Methods~\ref{sm:count}).
\end{table}}

\newcommand{\TabValidity}{%
\begin{table}[htbp]
\centering
\caption{REMHR as an index test against practice records (measured projects).}
\label{tab:s-validity}
\footnotesize
\begin{tabular}{@{}p{4.2cm}p{5.4cm}rrr@{}}
\toprule
Measure & Group & k/n & \% & 95\% CI \\
\midrule
sensitivity & FY2024 & 229/573 & 40.0 & 36.0 to 44.0 \\
specificity & FY2024 & 7,306/7,622 & 95.9 & 95.4 to 96.3 \\
PPV & FY2024 & 229/545 & 42.0 & 37.9 to 46.2 \\
sensitivity, unconditional & all continuing & 115/357 & 32.2 & 27.6 to 37.2 \\
specificity & all continuing & 4,966/5,145 & 96.5 & 96.0 to 97.0 \\
PPV & all continuing & 115/294 & 39.1 & 33.7 to 44.8 \\
sensitivity, unconditional & no racial or ethnic population named in FY2024 text & 10/188 & 5.3 & 2.9 to 9.5 \\
specificity & no racial or ethnic population named in FY2024 text & 4,820/4,888 & 98.6 & 98.2 to 98.9 \\
sensitivity, unconditional & revised, no longer named & 3/42 & 7.1 & 2.5 to 19.0 \\
specificity & revised, no longer named & 59/61 & 96.7 & 88.8 to 99.1 \\
sensitivity, unconditional & revised, still named & 36/46 & 78.3 & 64.4 to 87.7 \\
specificity & revised, still named & 18/39 & 46.2 & 31.6 to 61.4 \\
sensitivity, unconditional & unchanged & 66/81 & 81.5 & 71.7 to 88.4 \\
specificity & unchanged & 69/157 & 43.9 & 36.4 to 51.8 \\
retention & no racial or ethnic population named in FY2024 text & 10/17 & 58.8 & 36.0 to 78.4 \\
retention & revised, no longer named & 3/20 & 15.0 & 5.2 to 36.0 \\
retention & revised, still named & 35/38 & 92.1 & 79.2 to 97.3 \\
retention & unchanged & 66/67 & 98.5 & 92.0 to 99.7 \\
retention difference, unchanged minus no longer named & FY2025 &  & 83.5 & 61.5 to 93.3 \\
\bottomrule
\end{tabular}

\raggedright\footnotesize Reference standard: strict studied-population status for a racial or ethnic population. FY2025 groups are continuing projects by the fate of the summary. Retention conditions on carrying REMHR in FY2024. Wilson intervals; Newcombe interval for the difference. These describe agreement with the practice-record standard used here, not the classification accuracy of RCDC.
\end{table}}

\newcommand{\TabIpw}{%
\begin{table}[htbp]
\centering
\caption{Partial-verification correction of the studied-population gap.}
\label{tab:s-ipw}
\footnotesize
\begin{tabular}{@{}p{7.4cm}rrr@{}}
\toprule
Estimate & Value & 95\% CI & Pairs \\
\midrule
adjusted gap, measured pairs, unweighted & +8.4 & −5.1 to 21.1 & 332 \\
adjusted gap, inverse-probability-of-measurement weighted & +10.2 & −5.4 to 23.9 & 332 \\
share of pairs measured & 22.3\% & — & 1,488 \\
\bottomrule
\end{tabular}

\raggedright\footnotesize The unweighted gap is the main-text estimate. The share of pairs measured is a count over all revised signaled pairs, so it carries no interval. Probability of having a practice record fitted on all revised signaled pairs (activity, institute, mentions, density, title length, listed-term count, summary length); measured pairs weighted by its inverse \citep{begg1983}; text-adjusted linear probability model as in the main analysis; 2,000 bootstrap draws resampling projects.
\end{table}}

\newcommand{\TabTipping}{%
\begin{table}[htbp]
\centering
\caption{Studied-population gap corrected for non-differential misclassification of studied status.}
\label{tab:s-tipping}
\footnotesize
\begin{tabular}{@{}rrrrr@{}}
\toprule
Sensitivity & Specificity & Attenuation & Corrected gap & Upper 95\% limit \\
\midrule
0.5 & 0.90 & 0.44 & +19.1 & +47.9 \\
0.5 & 0.95 & 0.48 & +17.7 & +44.3 \\
0.5 & 1.00 & 0.50 & +16.7 & +41.8 \\
0.6 & 0.90 & 0.56 & +15.0 & +37.6 \\
0.6 & 0.95 & 0.62 & +13.5 & +34.0 \\
0.6 & 1.00 & 0.67 & +12.5 & +31.5 \\
0.7 & 0.90 & 0.64 & +13.1 & +32.9 \\
0.7 & 0.95 & 0.72 & +11.7 & +29.3 \\
0.7 & 1.00 & 0.79 & +10.7 & +26.8 \\
0.8 & 0.90 & 0.70 & +12.0 & +30.2 \\
0.8 & 0.95 & 0.79 & +10.6 & +26.6 \\
0.8 & 1.00 & 0.88 & +9.6 & +24.1 \\
0.9 & 0.90 & 0.74 & +11.3 & +28.4 \\
0.9 & 0.95 & 0.85 & +9.9 & +24.8 \\
0.9 & 1.00 & 0.94 & +8.9 & +22.3 \\
1.0 & 0.90 & 0.78 & +10.8 & +27.2 \\
1.0 & 0.95 & 0.89 & +9.4 & +23.6 \\
1.0 & 1.00 & 1.00 & +8.4 & +21.1 \\
\bottomrule
\end{tabular}

\raggedright\footnotesize Observed adjusted gap and its upper 95\% limit divided by the attenuation factor (Se + Sp − 1)π(1 − π)/[π*(1 − π*)], with π* the observed share studied. The minimum detectable effect is 19 points. Covariate adjustment can make attenuation larger than this factor.
\end{table}}

\newcommand{\TabLca}{%
\begin{table}[htbp]
\centering
\caption{Two-class latent class model for the three components of the reference standard.}
\label{tab:s-lca}
\footnotesize
\begin{tabular}{@{}lrr@{}}
\toprule
Indicator & Sensitivity & Specificity \\
\midrule
mesh & 0.21 & 0.99 \\
eligibility & 0.36 & 0.97 \\
enrollment & 0.96 & 0.83 \\
any of three (strict rule) & 0.98 & 0.79 \\
\bottomrule
\end{tabular}

\raggedright\footnotesize Black and Hispanic project–population pairs among measured projects where all three components are observable (1,532 pairs); latent prevalence 0.107. Assumes conditional independence of the three records given the latent class, which curated MeSH and trial records only approximate.
\end{table}}

\newcommand{\TabAuc}{%
\begin{table}[htbp]
\centering
\caption{Out-of-fold AUCs (5 folds grouped by project; 2,000 bootstrap draws resampling projects).}
\label{tab:s-auc}
\footnotesize
\begin{tabular}{@{}p{1.8cm}p{3.6cm}p{4.4cm}rrr@{}}
\toprule
Outcome & Sample & Model & Units & AUC & 95\% CI \\
\midrule
name survival & all revised, signaled & text only & 1,488 & 0.692 & 0.664 to 0.722 \\
name survival & measured subset, signaled & text only & 332 & 0.664 & 0.602 to 0.728 \\
name survival & measured subset, signaled & text + studied population & 332 & 0.666 & 0.602 to 0.729 \\
name survival & measured subset, signaled & studied population + population named only & 332 & 0.577 & 0.513 to 0.647 \\
name survival & measured subset, signaled & difference, text + studied population minus text only & 332 & 0.002 & −0.010 to 0.013 \\
REMHR lost & all continuing, REMHR in FY2024 & text only & 1,979 & 0.951 & 0.939 to 0.961 \\
REMHR lost & measured subset, REMHR in FY2024 & text only & 364 & 0.938 & 0.897 to 0.973 \\
REMHR lost & measured subset, REMHR in FY2024 & text + studied population & 364 & 0.936 & 0.894 to 0.972 \\
REMHR lost & measured subset, REMHR in FY2024 & difference, text + studied population minus text only & 364 & −0.002 & −0.010 to 0.005 \\
\bottomrule
\end{tabular}

\raggedright\footnotesize Calibration of the text-only survival model: calibration-in-the-large 0.003, slope 0.85. The fitted model (predict\_model.json; 15 features, L2 logistic, C = 1.0) is in the repository. Name-survival features: population named in the FY2024 title; log mentions; mention density; share of mentions sharing a sentence with a listed term, and whether all do; share of mentions in sentences stating aims or methods; log count of listed terms; title length; listed terms in the title; a listed form of the name; and indicators for the population named (Black, American Indian or Alaska Native, Asian, SGM, low income; Hispanic as reference). REMHR-loss features: whether the summary was revised; share of FY2024 sentences without a match in FY2025; whether a racial or ethnic population was named in FY2024 and whether it was no longer named; log counts of listed terms in FY2024 and FY2025.
\end{table}}

\newcommand{\TabTermDefs}{%
\begin{table}[htbp]
\centering
\caption{Share recorded as terminated under five definitions (\%).}
\label{tab:s-term-defs}
\footnotesize
\begin{tabular}{@{}lrrrrrr@{}}
\toprule
Population named & Awards & Ever & Still & HHS-reported & Excl. institution-wide & Non-training \\
\midrule
Sexual and gender minorities & 572 & 46.7 & 10.7 & 45.8 & 45.0 & 51.4 \\
Hispanic or Latino & 1,601 & 14.4 & 5.1 & 13.3 & 13.1 & 12.2 \\
Black or African American & 1,438 & 10.6 & 4.7 & 9.9 & 9.5 & 8.3 \\
American Indian or Alaska Native & 375 & 12.8 & 4.3 & 11.7 & 12.6 & 7.2 \\
Asian & 105 & 9.5 & 1.0 & 9.5 & 7.7 & 9.3 \\
Low income & 1,024 & 9.7 & 4.2 & 8.1 & 8.8 & 7.6 \\
Students & 2,990 & 13.9 & 6.5 & 11.8 & 11.9 & 6.7 \\
Health personnel & 1,798 & 6.1 & 1.8 & 4.4 & 3.8 & 5.5 \\
All awards & 55,302 & 4.8 & 1.7 & 3.2 & 3.0 & 3.7 \\
\bottomrule
\end{tabular}
\end{table}}

\newcommand{\TabTermAdj}{%
\begin{table}[htbp]
\centering
\caption{Adjusted differences in the share recorded as terminated, with the screen-term control.}
\label{tab:s-term-adj}
\footnotesize
\begin{tabular}{@{}lrrrr@{}}
\toprule
Population & \shortstack{Senate-list\\control} & \shortstack{95\% CI, institutions\\(2,000 draws)} & \shortstack{Screen-term\\control} & \shortstack{Crude\\RR} \\
\midrule
Sexual and gender minorities & +37.8 & 33.1 to 42.7 & +32.3 & 10.7 \\
Hispanic or Latino & +3.2 & 1.6 to 4.9 & −0.4 & 3.2 \\
Black or African American & +0.2 & −1.2 to 1.8 & −1.5 & 2.3 \\
American Indian or Alaska Native & +0.8 & −2.7 to 4.6 & −0.7 & 2.7 \\
Low income & −0.5 & −2.1 to 1.1 & −2.2 & 2.1 \\
Students & +2.3 & 1.0 to 3.6 & +0.8 & 3.2 \\
Health personnel & +0.2 & −1.4 to 2.2 & +0.3 & 1.3 \\
Caregivers & −1.5 & −2.7 to −0.2 & −2.1 & 0.8 \\
Veterans & −3.3 & −5.3 to −1.3 & −3.5 & 0.4 \\
\bottomrule
\end{tabular}

\raggedright\footnotesize Percentage points. Crude RR, against awards not naming the population. The screen-term control is circular, because the screen was reportedly built from terminated grants.
\end{table}}

\newcommand{\TabTermTopic}{%
\begin{table}[htbp]
\centering
\caption{Adjusted differences in the share recorded as terminated, with research-topic controls.}
\label{tab:s-term-topic}
\footnotesize
\begin{tabular}{@{}p{4.0cm}p{1.8cm}lp{2.4cm}rrr@{}}
\toprule
Specification & Awards & Outcome & Population & Named & Difference & 95\% CI \\
\midrule
main: log listed terms, institute, activity & all awards & ever & Sexual and gender minorities & 572 & +37.8 & 33.1 to 42.7 \\
main: log listed terms, institute, activity & all awards & ever & Hispanic or Latino & 1,601 & +3.2 & 1.6 to 4.9 \\
main: log listed terms, institute, activity & all awards & ever & Black or African American & 1,438 & +0.2 & −1.2 to 1.8 \\
plus 50 RCDC topic categories & all awards & ever & Sexual and gender minorities & 572 & +37.4 & 32.6 to 42.1 \\
plus 50 RCDC topic categories & all awards & ever & Hispanic or Latino & 1,601 & +2.5 & 0.7 to 4.4 \\
plus 50 RCDC topic categories & all awards & ever & Black or African American & 1,438 & −1.0 & −2.6 to 0.7 \\
plus 50 RCDC topic categories & all awards & still & Sexual and gender minorities & 572 & +7.0 & 3.9 to 10.2 \\
plus 50 RCDC topic categories & all awards & still & Hispanic or Latino & 1,601 & +0.5 & −0.7 to 1.8 \\
plus 50 RCDC topic categories & all awards & still & Black or African American & 1,438 & +0.7 & −0.5 to 2.0 \\
plus topics, listed terms split (population names / other) & all awards & ever & Sexual and gender minorities & 572 & +31.9 & 26.7 to 37.2 \\
plus topics, listed terms split (population names / other) & all awards & ever & Hispanic or Latino & 1,601 & +2.1 & 0.2 to 3.9 \\
plus topics, listed terms split (population names / other) & all awards & ever & Black or African American & 1,438 & −3.7 & −5.5 to −2.0 \\
plus 50 RCDC topic categories & HIV/AIDS awards & ever & Sexual and gender minorities & 262 & +31.8 & 25.1 to 38.6 \\
plus 50 RCDC topic categories & other awards & ever & Sexual and gender minorities & 310 & +41.2 & 35.4 to 47.1 \\
\bottomrule
\end{tabular}

\raggedright\footnotesize Percentage points. Linear probability models; all specifications hold the institute and the activity code fixed. Topic categories are indicators for the 50 most common RCDC categories of the FY2024 award. Population-name terms are Senate-list terms inside the population patterns; other terms are the rest. Ever, recorded as terminated at any point; still, still recorded as terminated on March 23, 2026. Intervals resample institutions (2,000 draws).
\end{table}}

\newcommand{\TabPractice}{%
\begin{table}[htbp]
\centering
\caption{Trial changes for the population named, among trials that could show the change.}
\label{tab:s-practice}
\footnotesize
\begin{tabular}{@{}p{3.0cm}p{4.4cm}rrrr@{}}
\toprule
Outcome & Name in the summary & Pairs & Events / at risk & \% & 95\% CI \\
\midrule
trial ended & Name removed & 118 & 8 / 179 & 4.5 & 2.3 to 8.6 \\
trial ended & Revised, still named & 115 & 1 / 141 & 0.7 & 0.1 to 3.9 \\
trial ended & Unchanged & 330 & 13 / 529 & 2.5 & 1.4 to 4.2 \\
trial ended & Difference, name removed minus unchanged &  &  & +2.0 & −1.4 to 5.8 \\
trial ended & Difference, name removed minus revised, still named &  &  & +3.8 & 0.6 to 7.2 \\
eligibility dropped the population & Name removed & 38 & 7 / 45 & 15.6 & 7.7 to 28.8 \\
eligibility dropped the population & Revised, still named & 54 & 5 / 59 & 8.5 & 3.7 to 18.4 \\
eligibility dropped the population & Unchanged & 96 & 5 / 105 & 4.8 & 2.1 to 10.7 \\
eligibility dropped the population & Difference, name removed minus unchanged &  &  & +10.8 & −0.5 to 24.7 \\
eligibility dropped the population & Difference, name removed minus revised, still named &  &  & +7.1 & −6.1 to 23.3 \\
new trial naming the population & Name removed & 204 & 13 / 204 & 6.4 & 3.8 to 10.6 \\
new trial naming the population & Revised, still named & 177 & 22 / 177 & 12.4 & 8.4 to 18.1 \\
new trial naming the population & Unchanged & 452 & 28 / 452 & 6.2 & 4.3 to 8.8 \\
new trial naming the population & Difference, name removed minus unchanged &  &  & +0.2 & −3.9 to 4.5 \\
new trial naming the population & Difference, name removed minus revised, still named &  &  & −6.1 & −12.1 to −0.2 \\
\bottomrule
\end{tabular}

\raggedright\footnotesize Pairs are a continuing award and a signaled population named in its FY2024 summary. Trial ended: a linked trial active in January 2025 (recruiting, active and not recruiting, or enrolling by invitation) that was terminated, suspended or withdrawn by September 2026; at risk, trial links active in January 2025 and registered in both snapshots. Eligibility dropped the population: a linked trial whose January 2025 criteria named the population and whose September 2026 criteria did not; at risk, trial links naming it in January 2025. New trial: any linked trial first registered from January 2025 whose criteria name the population, per pair. Group rates with Wilson intervals; differences in percentage points, with intervals resampling projects (2,000 draws).
\end{table}}

\newcommand{\TabRecheck}{%
\begin{table}[htbp]
\centering
\caption{Paired review of the FY2024 and FY2025 summaries: labels after arbitration.}
\label{tab:s-recheck}
\footnotesize
\begin{tabular}{@{}p{3.6cm}lrrrr@{}}
\toprule
Question & Label & \shortstack{Strict,\\name removed} & \shortstack{Random 50,\\name removed} & \shortstack{Strict, still\\named (control)} & κ \\
\midrule
FY2025 reference to the population & broader & 2 & 5 & 0 & 0.71 \\
 & named & 1 & 1 & 10 &  \\
 & none & 16 & 44 & 0 &  \\
 & specific & 1 & 0 & 0 &  \\
Research described & recast & 14 & 35 & 3 & 0.47 \\
 & rewritten & 1 & 0 & 1 &  \\
 & same & 5 & 15 & 6 &  \\
FY2025 summary alone meets the definition & no & 17 & 46 & 2 & 0.78 \\
 & unclear & 1 & 3 & 1 &  \\
 & yes & 2 & 1 & 7 &  \\
Automated detection of the group & correct & 19 & 44 & 9 & 0.53 \\
 & partly & 1 & 5 & 1 &  \\
 & wrong & 0 & 1 & 0 &  \\
\bottomrule
\end{tabular}

\raggedright\footnotesize Items: 20, 50 and 10. Labels are those of reviewers 3 and 2 where they agreed and of reviewer 1 where they differed (36 of 80 items). κ, Cohen's kappa between reviewers 3 and 2 before arbitration, on items both answered. The other questions were answered only for groups the reviewers found named in FY2024.
\end{table}}

\newcommand{\TabRecheckTiming}{%
\begin{table}[htbp]
\centering
\caption{When the qualifying 2025–2026 work was done: labels after arbitration.}
\label{tab:s-recheck-timing}
\footnotesize
\begin{tabular}{@{}p{5.4cm}rrrrr@{}}
\toprule
Items (source review) & \shortstack{Article dates\\work 2025+} & \shortstack{Trial listed\\as active} & \shortstack{Work before\\2025} & Undated & None \\
\midrule
Strict, lost REMHR, both reviewers E & 0 & 5 & 1 & 1 & 0 \\
Strict, lost REMHR, one reviewer E & 0 & 4 & 0 & 0 & 0 \\
Strict, controls & 2 & 7 & 2 & 1 & 0 \\
Random 50, arbitrated E & 0 & 0 & 8 & 1 & 0 \\
Random 50, one reviewer E & 0 & 0 & 1 & 0 & 1 \\
Random 50, controls & 0 & 0 & 3 & 0 & 0 \\
\bottomrule
\end{tabular}

\raggedright\footnotesize Items judged by at least one reviewer, in the current-eligibility review or the random sample, to show 2025–2026 research meeting the definition (37). E, research meeting the definition in 2025–2026. Labels as in the previous table (13 of 37 items arbitrated); κ for timing 0.74.
\end{table}}

\newcommand{\TabTrialActivity}{%
\begin{table}[htbp]
\centering
\caption{Registry records of the trials behind the evidence of ongoing work.}
\label{tab:s-trial-activity}
\footnotesize
\begin{tabular}{@{}p{3.6cm}rrrrrr@{}}
\toprule
Projects (timing review) & Projects & \shortstack{With a\\qualifying trial} & \shortstack{Update posted\\2025+} & \shortstack{Status verified\\2025+} & \shortstack{Active on\\Sep. 30, 2026} & \shortstack{Started\\2025+} \\
\midrule
Strict, lost REMHR, both reviewers E & 5 & 4 & 4 & 4 & 4 & 2 \\
Strict, lost REMHR, one reviewer E & 4 & 4 & 4 & 4 & 4 & 2 \\
Strict, controls & 7 & 7 & 6 & 6 & 7 & 2 \\
\bottomrule
\end{tabular}

\raggedright\footnotesize Projects whose evidence of ongoing work was a linked trial listed as recruiting, enrolling by invitation or active in September 2026 (timing review, labels after arbitration). A qualifying trial has that status in the September 2026 AACT snapshot and eligibility criteria naming a population the project concerns. Dates come from the current ClinicalTrials.gov record (September 30, 2026). A project counts when any qualifying trial meets the column's condition.
\end{table}}

\newcommand{\TabNamesReturn}{%
\begin{table}[htbp]
\centering
\caption{FY2026 records of FY2024 REMHR continuations, by what happened to the name in FY2025.}
\label{tab:s-names-return}
\footnotesize
\begin{tabular}{@{}p{4.4cm}rrrr@{}}
\toprule
FY2025 summary & Projects & \shortstack{FY2026\\record} & \shortstack{Names the\\population} & \shortstack{Text as\\in FY2025} \\
\midrule
Name removed & 281 & 200 & 3 & 163 \\
Strict 17, lost REMHR & 17 & 12 & 0 & 11 \\
Revised, still named & 282 & 184 & 170 & 113 \\
Unchanged & 782 & 496 & 444 & 316 \\
\bottomrule
\end{tabular}

\raggedright\footnotesize Continuing projects that carried REMHR in FY2024 with an FY2025 non-competing record; the strict 17 are a subset of the first row. Groups include projects whose FY2024 entry gave a project end before January 20, 2025, which Figure~4A of the article excludes. FY2026 records from NIH RePORTER on September 30, 2026, before the FY2026 ExPORTER file and FY2026 RCDC categories were published. Text as in FY2025: title, abstract and public health relevance unchanged; formatting differences between RePORTER and the ExPORTER files make this count a lower bound.
\end{table}}

\newcommand{\TabDecomposition}{%
\begin{table}[htbp]
\centering
\caption{Accounting decomposition of the FY2024–FY2025 change in REMHR total award costs.}
\label{tab:s-decomposition}
\footnotesize
\begin{tabular}{@{}p{9.6cm}rr@{}}
\toprule
Component & Projects & \$ million \\
\midrule
Kept REMHR in both years: change in cost & 1,530 & −39.2 \\
Lost REMHR, still funded: summary no longer names a racial or ethnic population & 237 & −125.0 \\
Lost REMHR, still funded: summary revised, population still named & 35 & −23.3 \\
Lost REMHR, still funded: summary unchanged & 5 & −3.2 \\
Lost REMHR, still funded: no racial or ethnic population named in FY2024 & 187 & −125.0 \\
Absent in FY2025: recorded as terminated & 150 & −71.7 \\
Absent in FY2025: not recorded as terminated & 838 & −505.1 \\
REMHR in FY2025 only: gained the category & 34 & +26.5 \\
REMHR in FY2025 only: new to the analysis universe & 447 & +312.2 \\
\midrule Total FY2024 & 2,982 & 1,876.8 \\
Total FY2025 & 2,011 & 1,323.1 \\
Change &  & −553.7 \\
\bottomrule
\end{tabular}

\raggedright\footnotesize Analysis universe: the latest eligible entry per core project and fiscal year, all application types. Components sum to the change without remainder. Dollars are total award costs of projects carrying the category, not spending on minority health research. Terminations from Grant Witness.
\end{table}}

\newcommand{\TabNulls}{%
\begin{table}[htbp]
\centering
\caption{Cohort trend in the informativeness of population names.}
\label{tab:s-nulls}
\footnotesize
\begin{tabular}{@{}p{7.2cm}rp{3.2cm}r@{}}
\toprule
Series & Estimate & 95\% interval & Units \\
\midrule
Informativeness trend, curated MeSH lift per cohort year (signaled) & −0.9 & −1.9 to 0.0 & 126,762 new awards \\
Informativeness trend, curated MeSH lift per cohort year (comparison) & −0.4 & −1.2 to 0.4 & 126,762 new awards \\
Informativeness trend, curated MeSH lift per cohort year (mechanical) & +0.5 & −0.2 to 1.2 & 126,762 new awards \\
Informativeness trend, curated MeSH lift per cohort year (difference signaled minus comparison) & −0.4 & −1.8 to 0.8 & 126,762 new awards \\
\bottomrule
\end{tabular}

\raggedright\footnotesize Percentage points per cohort year. The trend is null for signaled and comparison populations alike.
\end{table}}

\newcommand{\TabSentences}{%
\begin{table}[htbp]
\centering
\caption{Fate of FY2024 sentences naming a population in revised continuations.}
\label{tab:s-sentences}
\footnotesize
\begin{tabular}{@{}llp{4.4cm}rrr@{}}
\toprule
Population & Sentence holds & Outcome & Sentences & \% & 95\% CI \\
\midrule
Comparison & no listed term & sentence gone & 3,098 & 12.1 & 10.2 to 13.9 \\
Comparison & no listed term & sentence kept, name removed & 3,098 & 1.5 & 1.0 to 1.9 \\
Comparison & no listed term & sentence kept, name kept & 3,098 & 86.4 & 84.5 to 88.3 \\
Comparison & listed term & sentence gone & 1,179 & 33.5 & 30.3 to 36.9 \\
Comparison & listed term & sentence kept, name removed & 1,179 & 4.4 & 3.3 to 5.6 \\
Comparison & listed term & sentence kept, name kept & 1,179 & 62.1 & 58.6 to 65.5 \\
Signaled & no listed term & sentence gone & 2,201 & 29.4 & 26.4 to 32.7 \\
Signaled & no listed term & sentence kept, name removed & 2,201 & 31.9 & 28.7 to 35.0 \\
Signaled & no listed term & sentence kept, name kept & 2,201 & 38.7 & 35.1 to 42.7 \\
Signaled & listed term & sentence gone & 1,832 & 39.7 & 36.5 to 42.9 \\
Signaled & listed term & sentence kept, name removed & 1,832 & 28.6 & 25.7 to 31.4 \\
Signaled & listed term & sentence kept, name kept & 1,832 & 31.7 & 28.3 to 35.1 \\
\bottomrule
\end{tabular}

\raggedright\footnotesize Each sentence is matched to its most similar FY2025 sentence (token Jaccard); gone if no match reaches 0.5, otherwise kept, with the name kept or removed according to the matched sentence. Intervals resample awards (1,000 draws).
\end{table}}

\newcommand{\TabRecords}{%
\begin{table}[htbp]
\centering
\caption{REMHR in the project files used here against the published RCDC totals (\$ millions).}
\label{tab:s-records}
\footnotesize
\begin{tabular}{@{}lrrrrrrr@{}}
\toprule
Year & File date & Records & All records & Analysis universe & Published & \shortstack{Share,\\all records} & \shortstack{Share,\\universe} \\
\midrule
FY2023 & 2026-09-25 & 4,391 & 2,068 & 1,719 & 2,464 & 83.9 & 69.8 \\
FY2024 & 2026-09-25 & 4,615 & 2,184 & 1,877 & 2,612 & 83.6 & 71.9 \\
FY2025 & 2026-09-25 & 3,097 & 1,604 & 1,323 & 1,944 & 82.5 & 68.1 \\
\bottomrule
\end{tabular}

\raggedright\footnotesize Published: NIH RePORT, Estimates of Funding for Various Research, Condition, and Disease Categories, published June 23, 2026, accessed September 27, 2026. Shares are of the published total, for all records in the project files and for the analysis universe (NIH-administered extramural grants with a cost, excluding subprojects and supplements, latest entry per core project). 
\end{table}}

\newcommand{\TabForms}{%
\begin{table}[htbp]
\centering
\caption{Difference between the Hispanic and SGM listed-minus-unlisted loss gaps.}
\label{tab:s-forms}
\footnotesize
\begin{tabular}{@{}lrr@{}}
\toprule
Contrast & Points & 95\% CI \\
\midrule
Hispanic gap minus SGM gap & +5.4 & −4.1 to 14.8 \\
\bottomrule
\end{tabular}

\raggedright\footnotesize Normal approximation treating the two bootstrap intervals as independent; awards naming both Hispanic and SGM populations make this approximate.
\end{table}}

\newcommand{\TabSentenceCheck}{%
\begin{table}[htbp]
\centering
\caption{Human check of the automated sentence match: automated outcome by reviewers' judgment.}
\label{tab:s-sentence-check}
\footnotesize
\begin{tabular}{@{}llrrr@{}}
\toprule
Population & Automated outcome & Gone & \shortstack{Kept, name\\removed} & \shortstack{Kept, name\\kept} \\
\midrule
Signaled & Gone & 5 & 11 & 4 \\
 & Kept, name removed & 0 & 19 & 1 \\
 & Kept, name kept & 0 & 0 & 20 \\
 & Weighted share (\%) & 8.5 (3.4 to 15.4) & 47.6 (39.3 to 54.7) & 43.8 (38.7 to 50.8) \\
 & Automated share (\%) & 34.1 & 30.4 & 35.5 \\
Comparison & Gone & 5 & 0 & 10 \\
 & Kept, name removed & 0 & 1 & 9 \\
 & Kept, name kept & 0 & 0 & 15 \\
 & Weighted share (\%) & 6.0 (2.4 to 10.8) & 0.2 (0.0 to 0.7) & 93.8 (89.0 to 97.6) \\
 & Automated share (\%) & 18.0 & 2.3 & 79.7 \\
\bottomrule
\end{tabular}

\raggedright\footnotesize Sentences; columns are the reviewers' judgment (agreed labels of reviewers 3 and 2, arbitrated by reviewer 1 for 6; a broader term counts as name removed). Weighted shares apply the sample to all 4,033 signaled and 4,277 comparison sentences by stratum, with 95\% intervals resampling items within strata (2,000 draws).
\end{table}}

\newcommand{\TabFrozen}{%
\begin{table}[htbp]
\centering
\caption{REMHR in the project files against NIH's published RCDC project lists.}
\label{tab:s-frozen}
\footnotesize
\begin{tabular}{@{}llrrrrr@{}}
\toprule
Year & Projects & Project files & NIH list & Both & Files only & List only \\
\midrule
FY2024 & All REMHR projects & 4,615 & 4,618 & 4,615 & 0 & 3 \\
FY2025 & All REMHR projects & 3,097 & 3,106 & 3,097 & 0 & 9 \\
FY2024 & Reviewed items, status agrees &  &  & 50 of 50 &  &  \\
FY2025 & Reviewed items, status agrees &  &  & 50 of 50 &  &  \\
\bottomrule
\end{tabular}

\raggedright\footnotesize NIH list: the published RCDC project listing for REMHR (report.nih.gov, category 335), downloaded September 27, 2026. Retention with the NIH lists: unchanged, 66 of 67; revised, still named, 35 of 38; revised, no longer named, 3 of 20, identical to the project files.
\end{table}}

\newcommand{\TabAuditCross}{%
\begin{table}[htbp]
\centering
\caption{Projects whose summaries lost the name: evidence of continued work by FY2025 category status.}
\label{tab:s-audit-cross}
\footnotesize
\begin{tabular}{@{}>{\raggedright\arraybackslash}p{2.0cm}p{4.2cm}rrrr@{}}
\toprule
Source & Evidence & \shortstack{Lost\\REMHR} & \shortstack{FY2025\\cost} & \shortstack{Kept\\REMHR} & \shortstack{FY2025\\cost} \\
\midrule
Reviewer 1 & Later records show research on the population & 12 & \$8.1M & 3 & \$1.8M \\
 & Historical only & 1 & s & 0 & — \\
 & Unknown & 4 & \$1.9M & 0 & — \\
Both reviewers & Later records show research on the population & 10 & \$6.4M & 3 & \$1.8M \\
 & Historical only & 1 & s & 0 & — \\
 & Unknown & 1 & s & 0 & — \\
 & Reviewers disagree & 5 & — & 0 & — \\
Record evidence & Trial recruiting, active or enrolling by invitation in September 2026, naming the population & 8 & \$5.0M & 3 & \$1.8M \\
 & 2025–2026 article only & 3 & \$2.3M & 0 & — \\
 & Neither & 6 & \$3.2M & 0 & — \\
Overlap of the two checks & Both checks & 7 & \$4.6M & 3 & \$1.8M \\
 & Reviewers' check only & 3 & \$1.9M & 0 & — \\
 & Trial check only & 1 & s & 0 & — \\
 & Neither & 6 & \$3.6M & 0 & — \\
\bottomrule
\end{tabular}

\raggedright\footnotesize The 20 continuing projects that strictly study a racial or ethnic population, carried REMHR in FY2024 and whose summaries no longer named the population. Both reviewers: items given the same label by reviewers 1 and 2. Overlap: the reviewers' check is both reviewers judging 2025–2026 work verified; the trial check is the record rule above. Record evidence is read from the records, not the reviewers: a linked trial listed as recruiting, active or enrolling by invitation in its September 2026 registration record whose eligibility names the population (a trial not yet recruiting does not count); otherwise a linked article of 2025–2026 whose title or abstract names it. Costs rounded to \$0.1M; s, suppressed for cells with fewer than 3 projects.
\end{table}}

\newcommand{\TabThresholds}{%
\begin{table}[htbp]
\centering
\caption{Fate of sentences naming a population under three matching thresholds (\%).}
\label{tab:s-thresholds}
\footnotesize
\begin{tabular}{@{}lrrrrrr@{}}
\toprule
Jaccard & \shortstack{Signaled\\gone} & \shortstack{name\\removed} & \shortstack{name\\kept} & \shortstack{Comparison\\gone} & \shortstack{name\\removed} & \shortstack{name\\kept} \\
\midrule
0.4 & 27.7 & 35.5 & 36.7 & 15.1 & 3.1 & 81.8 \\
0.5 & 34.1 & 30.4 & 35.5 & 18.0 & 2.3 & 79.7 \\
0.6 & 42.0 & 24.3 & 33.7 & 21.6 & 1.7 & 76.7 \\
\bottomrule
\end{tabular}

\raggedright\footnotesize FY2024 sentences in revised continuations (4,033 signaled, 4,277 comparison). The main analysis uses 0.5.
\end{table}}

\TabSentenceCheck
\TabAuditCross
\TabAuditCurrent
\TabRecheckTiming
\TabTrialActivity
\TabAuditExpand
\TabRecheck
\TabDecomposition
\TabFrozen
\TabRecords
\TabTermAdj
\TabTermTopic
\TabPractice
\TabNamesReturn
\TabNulls
\TabSentences
\TabThresholds
\TabForms
\TabIpw
\TabTipping
\TabLca
\TabAuc
\TabAudit
\TabValidity
\TabTermDefs
\TabEstimands

\end{document}